\documentclass{article}
\usepackage{ltwol2e}

\input{psfig}

\def\Journal#1#2#3#4{{#1} {\bf #2}, #3 (#4)}

\def\ANP{{\em Annals Phys.}}
\def\NPB{{\em Nucl. Phys.} B}
\def\PLB{{\em Phys. Lett.}  B}
\def\PRL{\em Phys. Rev. Lett.}
\def\PRD{{\em Phys. Rev.} D}
\def\ZPC{{\em Z. Phys.} C}

\def\be{\begin{equation}}
\def\ee{\end{equation}}
\def\bea{\begin{eqnarray}}
\def\eea{\end{eqnarray}}
\def\Dslash{{\rm D}\!\!\!\!/\,}

\def\chpt{\hbox{$\chi$PT}}
\def\Qchpt{\hbox{Q$\chi$PT}}
\def\dyn{{\rm dyn}}
\def\val{{\rm val}}
\def\NPA{{\em Nucl. Phys.} A}
\def\NPBPS{{\em Nucl. Phys.} B (Proc. Suppl.)}
\def\JHEP{{\em JHEP} }

\begin{document}

\title{PROGRESS IN LATTICE GAUGE THEORY}

\author{STEPHEN R. SHARPE}

\address{Physics Department, Box 351560, University of Washington,
Seattle, WA 98195-1560, USA \\E-mail: sharpe@phys.washington.edu}

\twocolumn[\maketitle\abstracts{ 
I summarize recent progress in lattice gauge theory,
with particular emphasis on results from numerical simulations.
A major success has been the determination of the light hadron spectrum 
in the quenched approximation
with sufficient accuracy to expose statistically significant disagreements
with the experimental spectrum. 
The differences are, however, fairly small, $\sim 5-10\%$.
The data are also accurate enough to show evidence for artifacts of
quenching predicted by chiral perturbation theory.
I give an update on results for light quark masses, the kaon $B$-parameter, 
and the decay constants and $B$-parameters of heavy-light mesons.
Most of these are known in the quenched approximation to 
$\sim10\%$ accuracy or better, and preliminary estimates of 
quenching errors are of comparable size.
One exception is the light quark masses, for which
the quenching errors appear to be larger.
I discuss the computational requirements for simulations of QCD with
all approximations controlled,
and argue that they will likely begin once computers sustain 
about 10 Teraflops.
This is 30-40 times faster than present state-of-the-art machines.
This estimate assumes that improvements in the discretization
of lattice fermions are sufficient to allow continuum extrapolations to
be made with a minimum lattice spacing of $\approx 0.1\,$fm.
I review results obtained with improved discretizations and conclude
that they satisfy this requirement in most cases.
Examples of successful improvement include the calculation of
the glueball spectrum and excited heavy-quark potentials in pure 
Yang-Mills theory.
Finally, I discuss recent developments which may allow simulations 
of QCD with full chiral symmetry even at finite lattice spacing.
}]

\section{Introduction}

In this talk I aim to give an overview of recent progress in 
lattice gauge theory,
to report on the status of calculations of various quantities 
of phenomenological interest,
and to give my expectations for future progress.
I mainly consider results from numerical simulations of QCD,
although towards the end I discuss some ``analytical'' advances.
Given constraints of time and space, I have been highly selective,
and I apologize in advance to those whose work I have not included.
An excellent source for more details is the proceedings of the
annual lattice conference,
the most recent of which was held in July 1998~\cite{lat98}.
Previous reviews in the ICHEP and Lepton-Photon series
(by L\"uscher last year~\cite{luscher97}, and by Flynn in 1996~\cite{flynn96})
have different emphases than mine and are also highly recommended.

The most important feature of the lattice is that
it provides a non-perturbative regularization of field theory.
This is necessary in order to investigate the properties of strongly
coupled theories. The prime example is QCD at low energies or temperatures,
and this is the major focus of numerical lattice simulations.
There is, however, another non-perturbative sector of the standard model,
namely the electroweak sector 
close to the finite temperature symmetry-restoration transition. 
The properties of this transition are being actively investigated 
using lattice methods. These simulations are simpler than those for QCD,
since fermions can, to a good approximation, be integrated out analytically. 
The status of calculations is described in the recent review 
of Laine and Rummukainen~\cite{lainelat98}, and I will not discuss them here.

The ultimate goals of simulations of QCD include the following:
\begin{itemize}
\item
To test that QCD is indeed
the correct theory of the strong interactions;
\item
To make predictions for exotic mesons and baryons; 
\item
To determine the host of hadronic matrix
elements which are needed to extract the underlying quark parameters
(masses and weak mixing angles) from experimental results.
\end{itemize}
These are the topics I discuss in this talk.
There are other important goals which I will not discuss:
\begin{itemize}
\item
The attempt to gain at least a qualitative understanding of the
physics of confinement and chiral symmetry breaking
(a recent review is that by Negele~\cite{negelelat98});
\item
To understand the behavior of QCD at finite temperature and density
(see reviews by Kanaya~\cite{kanayafiniteT} and 
Laermann~\cite{laermannlat97}).
\end{itemize}
Another interesting topic I will cover only in passing is the status of
attempts to discretize chiral gauge theories. For reviews see
Refs.~\cite{niedermayerlat98,herbertichep98,maartenlat97}.

The outline of this review is as follows. Following a lightning
description of lattice calculations, 
which serves also to introduce some common terminology,
I explain why simulations of lattice QCD are so computationally
intensive, and indicate how close we are to controlling all sources
of error. In sec.~\ref{sec:QQCD} I describe the status of
quenched calculations of the spectrum.
Section~\ref{sec:imp} gives an update on the successes of
the ``improvement program'' at reducing discretization errors,
successes that are illustrated using the examples of the
spectrum of glueballs and potentials in the $SU(3)$ Yang-Mills theory.
Section~\ref{sec:QCD} gives a brief update on the status of 
spectrum calculations including dynamical quarks.
I then give a summary of the results for those hadronic matrix elements
which are calculated most reliably.
Section~\ref{sec:chiral} gives an overview of the recent progress 
in formulating lattice QCD with exact chiral symmetry.
 
\section{Why are lattice QCD calculations hard?}
\label{sec:LQCD}

Numerical simulations of gauge theories have been underway since
Creutz's seminal work almost twenty years ago 
(see, e.g., Ref.~\cite{creutzsu2}).
The first simulations including fermions (in the quenched approximation)
followed soon thereafter~\cite{parisi}.
Yet we are still not at the point of having reliable results for QCD.
In this section I will describe why the simulations are so difficult,
and explain what it will take to get reliable results.

I begin with an quick review of lattice methodology.
Wilson taught us how to discretize QCD while maintaining gauge 
invariance~\cite{WilsonLQCD}:
fermion fields are placed on the sites of a lattice, 
while the gauge fields live on the intervening links.
The simplest lattice, used in most numerical simulations,
is a hypercubic array, with equal spacings in all directions. 
The symmetry between sites makes the system well suited for
numerical implementation on parallel computers.
It is advantageous for some applications to make one or more lattice
spacings different, and I will mention an example of this below.
For numerical simulations one must also truncate to finite volume.
What results is a finite dimensional quantum mechanical system that
can be simulated using Feynman's sum over paths.
This is only practical if the time direction is imaginary,
so that one is in effect studying 4-dimensional statistical mechanicals.

The fermionic path integral is over anticommuting variables,
and is usually done by hand\footnote{%
There is, however, a recent paper in which the fermion integral is done
numerically~\cite{creutzgrassman}.},
resulting in the partition function 
\begin{equation}
Z_{\rm QCD} = \int [dU] \exp(-S_g) \prod_q 
\det(\Dslash_{\rm lat}+ m_q^\dyn)
\,.
\label{eq:ZQCD}
\end{equation}
In other words, the path integral is over all gauge fields $U$ 
(one $SU(3)$ matrix per link), weighted by the gauge action $S_g$
and the effective action 
resulting from fermion loops.
$S_g$ is a discretization of the continuum gauge action, and is
commonly taken to be that proposed by Wilson~\cite{WilsonLQCD}.
This has discretization errors of $O(a^2)$.
$\Dslash_{\rm lat}$ is a discretization of the Dirac operator.
The most common choices are Wilson fermions~\cite{Wilsonquarks}
(or improvements thereof which have smaller discretization errors),
and staggered fermions~\cite{staggered}.
Unimproved Wilson fermions have discretization errors of $O(a)$,
whereas staggered fermions have errors of $O(a^2)$.

Physical quantities are obtained from correlation functions. 
The simplest examples are two-point functions such as 
that of the pseudoscalar density,
\begin{equation}
C_P(x) = -\langle{ \bar u\gamma_5 d(x) \ \bar d\gamma_5 u(0) }\rangle \,.
\end{equation}
From this one can extract the mass of the $\pi^{-}$ 
(and, less easily, of its excited states).
Integrating out the fermions analytically gives the following expression
\begin{eqnarray}
\lefteqn{C_P(x) = Z_{\rm QCD}^{-1} 
\int  [dU] \exp(-S_g) \prod_q \det(\Dslash_{\rm lat} + m_{q}^\dyn)} \nonumber\\
&&\times
{\sf Tr}\left[
\gamma_5 \left({1\over \Dslash_{\rm lat}+m_d^\val}\right)_{x,0} \gamma_5
\left({1\over \Dslash_{\rm lat}+m_u^\val}\right)_{0,x} \right] \,,
\label{eq:CPdef}
\end{eqnarray}
which is suitable for numerical evaluation.
Thus, to calculate a correlation function,
one has to calculate quark ``propagators'' 
(i.e. inverses of the discretized Dirac operator plus mass term)
in the background gluon configuration,
connect them together as shown by the trace, 
and average the result over gauge configurations weighted by the
total effective action.
The average is done using Monte-Carlo methods.

A very important distinction in the following is between ``dynamical''
and ``valence'' quark masses. The former are those which appear in the
determinant, while that latter occur in the propagators.
I have labeled the quark masses in Eq.~\ref{eq:CPdef} according to
this terminology.
In QCD, these two masses are the same, flavor by flavor, since they both
arise from the same original action.
But once one has integrated out the fermions, one has the option
of picking the masses to be different.
This is called the ``partially quenched'' approximation,
and is made in most large scale calculations at present.
Typically, the dynamical quark masses are heavier than the valence
quark masses. The fully quenched limit is that in which $m^\dyn\to\infty$,
in which case the fermion determinant is a constant and drops out of
the action. In summary, the partially quenched approximation means
taking $m^{\rm val} \ne m^{\rm dyn}$, and the limit $m^{\rm dyn}\to\infty$
with finite $m^{\rm val}$ is the quenched approximation.

It is important to realize that (partially) quenched theories 
are unphysical---for example, they do not have
an hermitian Hamiltonian.
These approximations are made simply out of necessity because of the
difficulty in simulating QCD with light dynamical masses.
To illustrate this difficulty I will use an approximate scaling law
for the number of floating-point operations (Flops) that are required
to generate a statistically independent gauge configuration 
using 2 flavors of dynamical staggered fermions and 
present algorithms~\cite{dougalg}
\begin{equation}
{\mbox{\# Flops} \over \mbox{Indep. Config.}} \approx
{6000 \times N_s^3 N_t \over (a m^\dyn)^{5/2}}
\label{eq:scalingI}
\end{equation}
Here $N_s$ and $N_t$ are the number of points in the spatial and Euclidean
time directions. Assuming a typical value for their ratio, $N_t/N_s=2$, this 
converts to
\begin{equation}
{\mbox{\# Tflop-hours} \over \mbox{100 Configs}}
\approx 11 f \left(L \over 2.5 {\rm fm}\right)^4 
	\left( 0.1 {\rm fm}\over a\right)^{6.5}
	\left( m_s/2 \over m^\dyn \right)^{2.5} \,,
\label{eq:scalingII}
\end{equation}
where $L=N_s a$ is the spatial size of the lattice in physical units,
and $m_s$ is the physical strange quark mass.
I choose 100 independent configurations in the denominator
since this is roughly the minimal number needed to obtain
small statistical errors in masses.
I have introduced a factor $f$ to account for differences in
simulation times between different discretizations of fermions.
Numerical examples of this formula are give in Table~\ref{tab:flops}.
Note that the lattice size is chosen to satisfy both
$L\ge 2.5\,$fm and $m_\pi L \ge 4.5$
to avoid significant finite volume effects.

\begin{table}[tb]
\begin{center}
\caption{Time for 100 QCD configurations on a 
Teraflop machine.}\label{tab:flops}
\vspace{0.2cm}
\begin{tabular}{|c|cc|ccc|} 
\hline 
\raisebox{0pt}[12pt][6pt]{$m^\dyn$} & 
\raisebox{0pt}[12pt][6pt]{$m_\pi \over m_\rho$ } & 
\raisebox{0pt}[12pt][6pt]{${L \over \mbox{fm}}$} & 
\raisebox{0pt}[12pt][6pt]{$a\!=\!.2\,$fm} &
\raisebox{0pt}[12pt][6pt]{$a\!=\!.1\,$fm} &
\raisebox{0pt}[12pt][6pt]{$a\!=\!.05\,$fm}\\
\hline
\raisebox{0pt}[12pt][6pt]{${m_s/ 2}$} & 
\raisebox{0pt}[12pt][6pt]{$0.56$ } & 
\raisebox{0pt}[12pt][6pt]{$2.5$} & 
\raisebox{0pt}[12pt][6pt]{$7.5 f \,$mins} &
\raisebox{0pt}[12pt][6pt]{$11 f \,$hrs} &
\raisebox{0pt}[12pt][6pt]{$42 f \,$days}\\
\hline
\raisebox{0pt}[12pt][6pt]{${m_s/8}$} & 
\raisebox{0pt}[12pt][6pt]{$0.3$ } & 
\raisebox{0pt}[12pt][6pt]{$3.5$} & 
\raisebox{0pt}[12pt][6pt]{$15 f \,$hrs} &
\raisebox{0pt}[12pt][6pt]{$58 f \,$days} &
\raisebox{0pt}[12pt][6pt]{$1.4 f \,$yrs}\\
\hline
\raisebox{0pt}[12pt][6pt]{${m_s/25}$} & 
\raisebox{0pt}[12pt][6pt]{$0.18$ } & 
\raisebox{0pt}[12pt][6pt]{$6.7$} & 
\raisebox{0pt}[12pt][6pt]{$0.4 f \,$yrs} &
\raisebox{0pt}[12pt][6pt]{$3.7 f \,$yrs} &
\raisebox{0pt}[12pt][6pt]{$330 f \,$yrs}\\
\hline
\end{tabular}
\end{center}
\end{table}
\vspace*{3pt}

The top row in the Table corresponds approximately to the
lightest quark masses used in present simulations of (unquenched) QCD.
Simulations at this quark mass are being done with the lattice spacings
down to $0.1\,$fm.
The bottom row corresponds to the physical average light-quark mass.
This is not quite QCD, since a dynamical strange quark is not included,
but it gives an indication of the CPU time required for a direct
simulation of QCD.
Finally, the middle row corresponds to a quark mass which should be light
enough that one can extrapolate from it to the physical values. 
In other words, we probably do not need to use masses
smaller than those of the middle row, because we can use chiral perturbation
theory to guide us in doing extrapolations. The chiral expansion parameter
is $m_\pi^2/m_\rho^2$, and this is $10\%$ for the middle row
(but $30\%$ for the top row).

A striking feature of these numbers
is how it becomes rapidly more difficult to simulate 
as the lattice spacing and quark mass decrease.
This follows directly from Eq.~\ref{eq:scalingII}.
Part of this increase is unavoidable (e.g. the number of lattice points
increases as $a$ decreases) but part is due to shortcomings of
present algorithms (e.g. the critical slowing down as $m_q\to0$).
Although I expect algorithms with improved scaling laws to be developed,
I use the scaling of present algorithms for my estimates. 

The dependence on $a$ makes clear the importance of the 
``improvement program'',
which aims to reduce discretization errors and thus allow simulations
at larger lattice spacings. In quenched simulations with unimproved
Wilson fermions, it turns out (as discussed below) that one needs to
go down to $a\approx 0.05\,$fm in order to extrapolate to the continuum limit.
It would be a great advance to be able to work with
a minimum lattice spacing of $a\approx 0.1\,$fm instead.
I will show later that this is likely to be possible using
improved Wilson fermions. With staggered fermions, on the other
hand, it appears that $a\approx 0.05\,$fm will be necessary.

Let us assume, then, that we work with improved Wilson fermions,
and can get away with a minimum $a$ of $0.1\,$fm,
and a minimum quark mass of $m_s/8$.
Then we need only consider the top left $2\times2$ block of the table.
What does this imply for the absolute time required?
To answer this I must discuss the factor $f$.
This represents the extra cost of doing simulations with improved 
Wilson fermions compared to staggered fermions.
Based on the experience of the CP-PACS collaboration~\cite{burkprivate}
this factor is at least $10$, implying that the simulation with
$m^{\rm dyn}=m_s/8$ and $a=0.1\,$fm would take almost 2 years on
a sustained Teraflop machine. To be realistic one should also include
a sizeable additional factor to account for 
the extra cost of including the dynamical strange quark,
the need to do multiple simulations in order to extrapolate, 
etc. Thus I expect that reliable results for
QCD itself will begin to emerge once we have 
machines sustaining 10 Teraflops.\footnote{%
This estimate is somewhat more pessimistic than that I gave in the talk.
Simulations with improved actions are somewhat slower than I had assumed.}
I want to stress that this is only 30-40 times 
faster than the present state-of-the-art machines.
These are the CP-PACS machine at the
University of Tsukuba, Japan, which peaks at $0.6\,$Tflops
and sustains $0.3\,$Tflops~\cite{iwasakimachine};
and the two Columbia-Riken/BNL machines~\cite{christmachine}
which together have a peak speed of 1 Tflop, 
and present efficiencies of about 25\%. 
CP-PACS has been in production mode for nearly 2 years,
while the Columbia machine ($0.4\,$Tflop peak) 
has been in production mode for 6 months,
and the Riken/BNL machine ($0.6\,$Tflop peak) 
is just becoming fully operational.

\begin{figure}
\center
\hspace{.2in}\psfig{figure=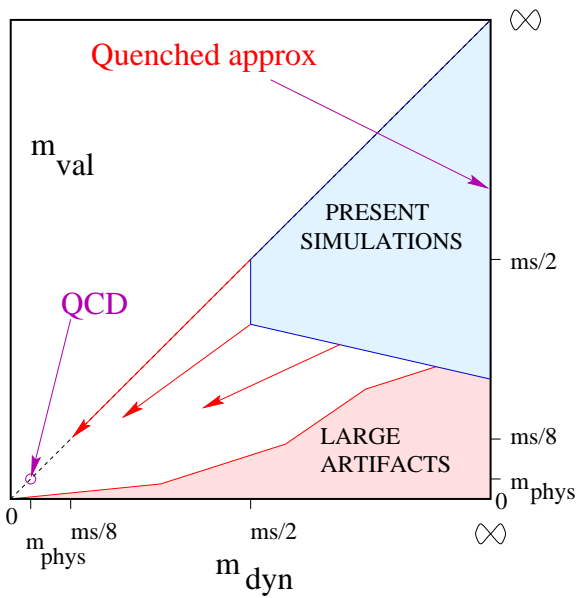,width=3.0in}
\vspace{-0.1in}
\caption{Approaching QCD.}
\label{fig:approach}
\end{figure}

Figure~\ref{fig:approach} gives a slightly different
view of the status of present simulations.
It shows a two-dimensional section through parameter space
(and thus hides the dependence on the lattice spacing and the
strange quark mass).
Theories along the diagonal are physical, 
while those off the diagonal are partially quenched and thus unphysical.
The fully quenched theory is on the righthand vertical axis at 
$m^\dyn=\infty$.
The upper shaded region roughly indicates where
present simulations are done.
Note that it is relatively inexpensive, for a given dynamical quark mass,
to map out a range of valence quark masses.
Thus progress will occur along the various arrows pointing towards
the physical point in the LH corner.
I wish to emphasize that, because of the ease of collecting partially
quenched data, it would be very useful if we could use the information from
these unphysical theories to aid the extrapolation to the physical theory.
This can perhaps be done using chiral perturbation theory,
and work in this direction has begun~\cite{ssPQ,GLPQ}.
There is, however, a limitation on how small one can take $m^\val$,
because certain artifacts of quenching diverge as $m^\val\to0$.
Examples of this in the quenched theory will be shown below.

To conclude this section, I list the
important issues that effect the progress towards simulations of
full QCD.
\bigskip

\begin{itemize}
\item
How large are quenching errors?
To what extent can we use quenched QCD as a guide to QCD?
\item
How small a lattice spacing is needed to extrapolate to the continuum?
By how much can we increase $a_{\rm min}$ with an improved action?
\item
How small a quark mass is needed to reliably extrapolate to QCD?
\item
Can we use (partially quenched) chiral perturbation theory to aid
extrapolation to the physical parameters?
\end{itemize}

\section{The spectrum of quenched QCD}
\label{sec:QQCD}

The CP-PACS collaboration presented this year final results from
their study of the spectrum of QCD in the quenched 
approximation~\cite{burkhalter,yoshie}.
Their aim is to control all errors, except that due to quenching,
and thus to obtain definitive results for the quenched spectrum.
They use:
\begin{itemize}
\item
the simplest discretizations of the gauge and fermion actions,
both due to Wilson. Thus the dominant errors are of $O(a)$, coming
from the fermions;
\item
a fixed physical volume large enough for finite size effects
to be negligible for single particle masses;
\item
four (small) lattice spacings to extrapolate to the continuum limit ($a=0$);
\item
five valence quark masses, spanning the range $m_\pi/m_\rho = 0.4-0.75$,
to extrapolate to the physical light quark masses;
\item
large ensembles of gauge configurations so that the final
errors, after extrapolation, are of size $1-3\%$.
\end{itemize}
Some parameters of their simulations are collected
in Table~\ref{tab:CPPACSparams}. I focus mainly on the CP-PACS results
for the spectrum since their computational power outstrips that
of other groups by an order of magnitude.

\begin{table}
\begin{center}
\caption{Parameters of CP-PACS quenched simulations.}\label{tab:CPPACSparams}
\vspace{0.2cm}
\begin{tabular}{|c|c|c|c|c|} 
\hline 
\raisebox{0pt}[12pt][6pt]{$\beta=6/g^2$} & 
\raisebox{0pt}[12pt][6pt]{$a$ (fm)} & 
\raisebox{0pt}[12pt][6pt]{size} & 
\raisebox{0pt}[12pt][6pt]{$L a$ (fm)} &
\raisebox{0pt}[12pt][6pt]{Configs.} \\
\hline
\raisebox{0pt}[12pt][6pt]{5.90} & 
\raisebox{0pt}[12pt][6pt]{$0.10$} & 
\raisebox{0pt}[12pt][6pt]{$32^3\times56$} & 
\raisebox{0pt}[12pt][6pt]{$3.3$} &
\raisebox{0pt}[12pt][6pt]{800} \\
\hline
\raisebox{0pt}[12pt][6pt]{6.10} & 
\raisebox{0pt}[12pt][6pt]{$0.078$} & 
\raisebox{0pt}[12pt][6pt]{$40^3\times70$} & 
\raisebox{0pt}[12pt][6pt]{$3.1$} &
\raisebox{0pt}[12pt][6pt]{600} \\
\hline
\raisebox{0pt}[12pt][6pt]{6.25} & 
\raisebox{0pt}[12pt][6pt]{$0.064$} & 
\raisebox{0pt}[12pt][6pt]{$48^3\times84$} & 
\raisebox{0pt}[12pt][6pt]{$3.1$} &
\raisebox{0pt}[12pt][6pt]{420} \\
\hline
\raisebox{0pt}[12pt][6pt]{6.47} & 
\raisebox{0pt}[12pt][6pt]{$0.05$} & 
\raisebox{0pt}[12pt][6pt]{$64^3\times112$} & 
\raisebox{0pt}[12pt][6pt]{$3.2$} &
\raisebox{0pt}[12pt][6pt]{150} \\
\hline
\end{tabular}
\end{center}
\end{table}
\vspace*{3pt}

\begin{figure}[t]
\vspace{-0.2in}
\psfig{figure=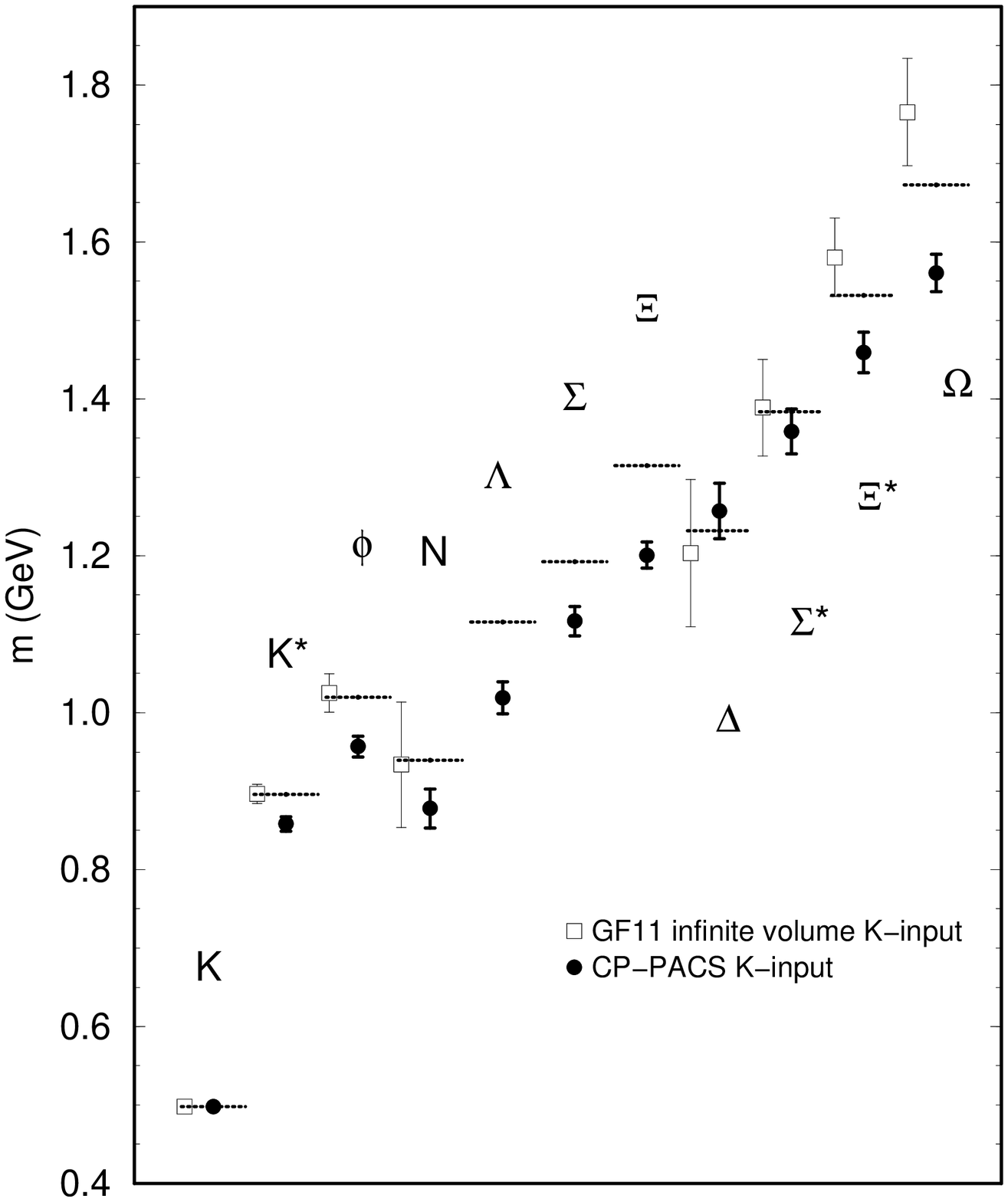,width=3.5in}
\vspace{-0.2in}
\caption{Quenched spectrum with $m_s$ fixed using $m_K$.}
\label{fig:CPPACS-K}
\end{figure}

\begin{figure}[t]
\vspace{-0.2in}
\psfig{figure=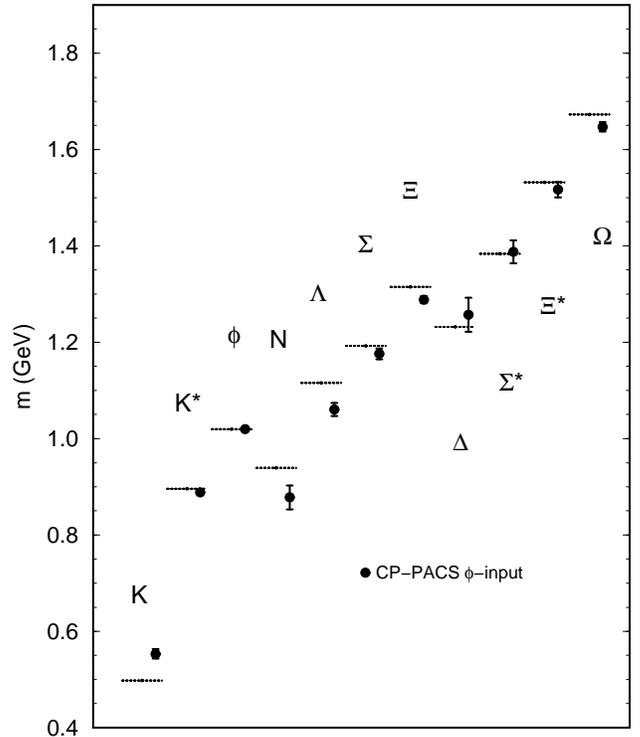,width=3.5in}
\vspace{-0.2in}
\caption{Quenched spectrum with $m_s$ fixed using $m_\phi$.}
\label{fig:CPPACS-phi}
\end{figure}

This calculation supersedes that done using the GF11 machine
and presented in 1994~\cite{GF11}.
The improvement, due to more than an order of magnitude increase
in computer power, is evident from the final results,
shown in Figs.~\ref{fig:CPPACS-K} and \ref{fig:CPPACS-phi}.
These figures give the masses of pseudoscalar and vector mesons
and octet and decuplet baryons. To understand them it is
useful to review how the parameters of lattice simulations are fixed
in quenched simulations.
First, for each lattice spacing one extrapolates in the average 
light quark mass to the value at which the physical $m_\pi/m_\rho$ ratio is 
obtained.\footnote{Note that isospin breaking effects are being ignored.}
The resulting rho mass is then used to determine the lattice spacing
(one extracts $m_\rho a$ from the lattice, so 
$a = (m_\rho a)/m_\rho^{\rm phys}$).
At this point all masses can be converted to units of GeV
(although one should keep in mind that one is really determining
the ratio of the masses to $m_\rho$).
Next, one interpolates in the lattice strange quark mass until a particular
strange particle has its physical mass---in the present cases the 
$K$ and $\phi$ are used.
And, finally, one extrapolates in the lattice spacing to $a=0$.
The results of this procedure are shown in the Figures---the pion and
rho masses are not shown since they are input parameters.

I draw two important lessons from these results.
First and foremost, the quenched spectrum disagrees with experiment!
This is an important check on the calculations, since we know that
we are not calculating in the correct theory.
The most significant disagreement is for the hyperfine splitting, 
$m_{K^*}-m_K$, which is $\sim\!10\%$ too small for either choice
of $m_s$. This is a $5-6\sigma$ discrepancy. 
The nucleon mass is 7\% too low (independent of $m_s$), 
although this is less significant ($2.5\sigma$).
Finally, the splittings within the octet and decuplet are too small, 
although this discrepancy is greatly reduced if $m_s$ is fixed
using $m_{\phi}$. However, the Gell-Mann--Okubo formula for octet baryons
and the equal-spacing rule for decuplet baryons
are both well satisfied for either value of $m_s$.

The second lesson is that quenched QCD yields a spectrum 
which is reasonably close ($5-10\%$) to that expected from QCD itself.
In other words, although quenching is a drastic approximation,
it gives us a model in which the gross properties of light hadrons
are numerically close to those in QCD.
This is important because we want to have an idea of the
accuracy of quenched results for other quantities,
such as the hadronic matrix elements discussed below.
One should keep in mind, however, that there is no universal
size to quenching errors.

Extrapolations play a key r\^ole in the CP-PACS analysis,
and I want to spend a little time discussing their reliability. 
I focus on the chiral extrapolation here, deferring discussion of
the continuum extrapolation until the next section.
To do the chiral extrapolation requires theoretical input 
as to the functional form, 
and this is provided by chiral perturbation theory (\chpt).
For the quenched theory one must use the quenched variant,
\Qchpt, and this yields the following predictions for the 
chiral expansions of pion and nucleon masses~\cite{BG,sschirallogs,LS}:
\begin{eqnarray}
m_\pi^2 &=& 2 A m_q 
\left(1 - \delta \log(2Am_q/\Lambda_\chi^2) + O(m_q) \right) 
\label{eq:Qchptmpi}\\
m_N &=& M_{0} - \delta C_1 m_\pi + C_2 m_\pi^2 + C_3 m_\pi^3 + 
O(m_\pi^4)
\label{eq:QchptmN}
\end{eqnarray}
In these expressions, $\delta$, $A$, $\Lambda_\chi$,
$M_0$ and the $C_i$ are constants which are related to the 
coefficients appearing in the quenched chiral Lagrangian.
The result for $m_\rho$ has the same form as that for $m_N$ 
but with different constants~\cite{boothvectors}.
The important features of these expressions are as follows.
\begin{itemize}
\item
Terms involving the constant $\delta$ are artifacts of quenching---they 
are not present in the chiral expansions in QCD.
They arise from loops containing the $\eta'$,
which, in the quenched theory, remains degenerate with the pion.
They give the dominant corrections to the leading order term 
as $m_q\to0$. In particular, the $\delta \log m_q$ term in
(\ref{eq:Qchptmpi}) diverges in the chiral limit, although the
contribution to $m_\pi^2$ itself vanishes in this limit.
\item
Terms not involving $\delta$ are present in both quenched and unquenched
theories, although the detailed expressions for the coefficients differ.
\item
Particle masses are predicted to be non-analytic functions of
the quark mass---the non-analyticity is
proportional to $m_q\log m_q$ in (\ref{eq:Qchptmpi}) and
to $m_\pi\sim m_q^{1/2}$ and $m_\pi^3\sim m_q^{3/2}$ in (\ref{eq:QchptmN}).\\
\end{itemize}

\noindent
In the past, most analyses have extrapolated using low order polynomials
in $m_q$, i.e. ignoring possible non-analytic terms.
These parameterizations have adequately represented the data.
The small statistical errors and long chiral level-arm 
of the CP-PACS calculation
make possible a test of the more complicated forms predicted \Qchpt.

\begin{figure}[tb]
\psfig{figure=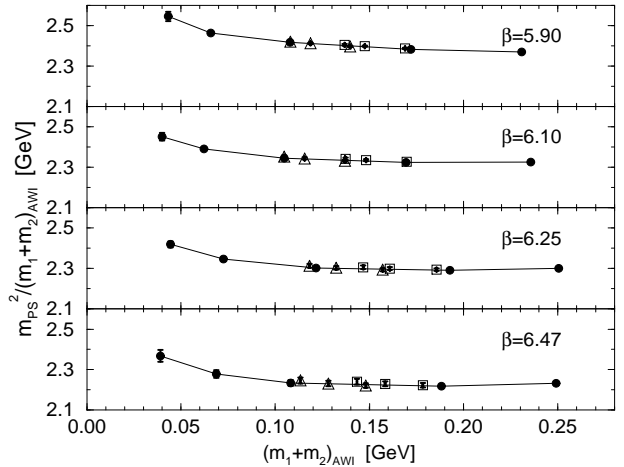,width=3.7in}
\caption{CP-PACS results for $y$ at all four lattice spacings.
Points with open triangles and squares are for non-degenerate quarks;
others are for degenerate quarks. The quark masses are in the 
$\overline{\rm MS}$ scheme at $\mu=2\,$GeV.
The lines are drawn to guide the eye and are not fits.}
\label{fig:CPPACSmpi2}
\end{figure}

The most dramatic effect predicted by \Qchpt\ is for the pion mass:
the quantity $y=m_\pi^2/(m_1+m_2)$ should diverge
logarithmically at small $m_q$. (See Ref.~\cite{BG} for the full
expression for non-degenerate quarks.) This is to be compared to QCD,
where $y$ linearly approaches a constant in the chiral limit.
Evidence for this effect has previously been seen with staggered
fermions (and is reviewed in Ref.~\cite{sslat96}), but not with Wilson
fermions. The CP-PACS results are shown in Fig.~\ref{fig:CPPACSmpi2}.
The predicted rise at low quark masses is seen for all four lattice spacings,
with the fits yielding consistent values of $\delta\sim0.10 \pm 0.02$.
This rise has also been seen by the FNAL group~\cite{thackerlat98},
who are able to work with very small quark masses by 
using a modification of the quenched approximation,
and by the QCDSF group using improved Wilson fermions~\cite{QCDSFdelta}.
The values for $\delta$ are all roughly consistent with each other, 
and with those obtained using other quantities~\cite{sslat96}.

The most economical explanation of these results is that
they are quenched artifacts due to $\eta'$ loops, as predicted by \Qchpt.
This interpretation is supported by the fact that fits to the
predicted chiral behavior of vector mesons, baryons
and pseudoscalar decay constants (including those 
for particles composed of non-degenerate quarks)
represent the CP-PACS data well~\cite{burkhalter,yoshie}.
It is reassuring that the lattice results are good enough to
pick out physical effects coming from loops of light particles,
and thus to make contact with the analytical predictions of \Qchpt.
It also gives credibility to attempts to use \Qchpt\ to 
estimate the size of quenching errors in other 
quantities~\cite{sslat96,SZfb,GLK2pi}.

On the other hand, one cannot rule out that the rise in $y$ is due 
to a collusion between terms analytic in $m_q$ of
linear, quadratic and cubic order.
CP-PACS have investigated this possibility by fitting all the masses
to polynomials of the lowest order required to give a good fit.
The resulting light hadron masses differ at finite lattice spacing
from those obtained using the \Qchpt\ forms, 
but it turns out that they agree within errors after extrapolation to
the continuum limit (see Fig.~\ref{fig:CPPACSa} below). 
In particular, the hyperfine splitting is still
significantly below the experimental value.
Thus the fact that quenched QCD disagrees with experiment appears
to be robust. 

The importance of understanding the chiral extrapolations has also been
stressed by the MILC collaboration,
who have done extensive studies using staggered fermions~\cite{MILCchiral}.

My overall conclusion concerning chiral fits from present data
is that, while it is clear that the lowest order polynomial fits
are inadequate once one goes below $m_\pi/m_\rho < 0.5$,
a complete resolution of the issue probably requires
smaller masses than those used by CP-PACS.
This is part of the reason why I chose $m_\pi/m_\rho=0.3$ 
(rather than the minimal value $m_\pi/m_\rho=0.4$ used by CP-PACS)
as the number to aim at in Table~\ref{tab:flops}.
Actually, sorting out the chiral extrapolation may be easier in the
unquenched theory, since there are no ``enhanced'' logarithms.
On the other hand, such logarithms are present in partially quenched
QCD~\cite{ssPQ,GLPQ} 
(and give rise to the artifacts in the lower shaded region in
Fig.~\ref{fig:approach}).

While the CP-PACS results are very impressive, they do rely on certain
assumptions about extrapolations.
It is clearly important to check them using at least one other
fermion discretization, for which the systematics will differ.
The MILC results with staggered fermions, 
while not as extensive (e.g. they do not consider
baryons composed on non-degenerate quarks), do provide a partial check.
They quote the extrapolated value $m_N/m_\rho=1.254\pm0.018\pm0.028$,
consistent with the experimental value $1.22$.
This is higher than the CP-PACS result of
$1.143 \pm 0.033$, but only by about $2.5 \sigma$.
It would be very interesting to see results from staggered fermions
for the $K-K^*$ hyperfine splitting.
There are also a number of large scale calculations underway
using improved Wilson fermions~\cite{QCDSFimp,QCDSFdelta,APETOVimp,APEimp},
but these do not yet have a range of lattice spacings or 
statistical accuracy comparable to that of the CP-PACS collaboration.

\section{The improvement program}
\label{sec:imp}

The discussion of sec.~\ref{sec:LQCD} highlighted the importance of reducing
discretization errors.
This is essential if we are to have unquenched simulations of QCD in the
near future, and it has been the focus of much research in the last
few years. In this section I report on the progress in this direction.

\begin{figure}[tb]
\psfig{figure=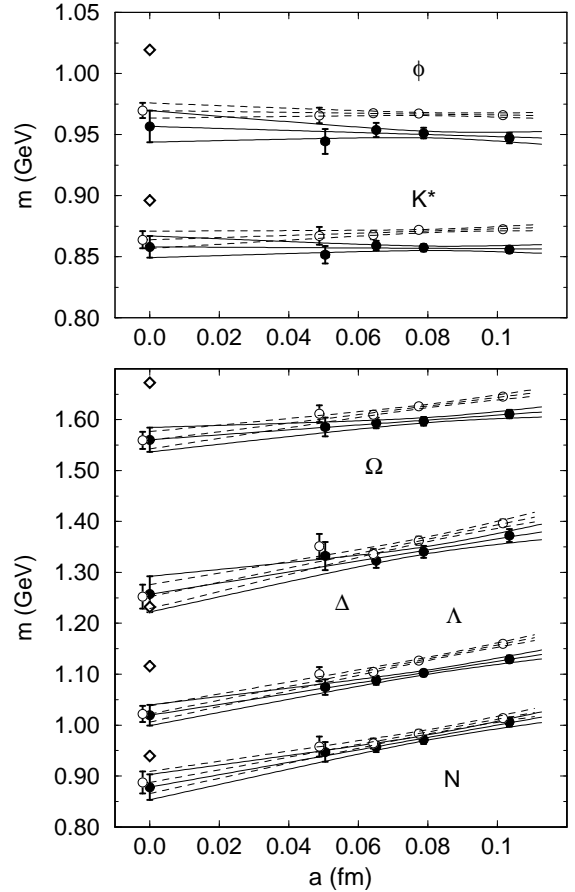,width=3.7in}
\caption{CP-PACS continuum extrapolations, with $m_s$ fixed using
$m_K$, and the scale in GeV set by $m_\rho$. 
Solid (open) points are from chiral extrapolations based 
on \Qchpt\ (polynomial) fits. 
Experimental values are shown by diamonds at $a=0$.}
\label{fig:CPPACSa}
\end{figure}

With an unimproved action, such as that for Wilson fermions,
the discretization errors start at $O(a)$
\begin{equation}
M_{\rm Had}^{\rm lat}(a) = M_{\rm Had}^{\rm cont} 
\left[ 1 + a \Lambda_1 + (a \Lambda_2)^2 + \cdots \right]
\label{eq:aextrap}
\end{equation}
The size of the corrections is set by non-perturbative physics,
so we expect $\Lambda_1\sim\Lambda_2 \sim \Lambda_{\rm QCD}$. 
Examples of continuum extrapolations from the CP-PACS data are
shown in Fig.~\ref{fig:CPPACSa}.
The fits are to a linear form ($\Lambda_2=0$), and give
values for $\Lambda_1$ in the range $0-300\,$MeV.
Note that a linear fit is reasonable for such small lattice spacings.
If one introduces a quadratic term of the expected size
the extrapolated values change by less than $1\sigma$.

It is useful to estimate the minimal lattice spacing ($a_{\rm min}$) 
needed to extrapolate to the continuum with a given accuracy.
I take the accuracy required after extrapolation to be $1\%$.
To approximate the extrapolation procedure I assume that
$M_{\rm had}$ and $dM_{\rm had}/da$ 
can be determined accurately  at $a_{\rm min}$, 
and used to extrapolate linearly to zero.
For the quadratic function in Eq.~\ref{eq:aextrap}
the resulting error is $-(a_{\rm min}\Lambda_2)^2$.
Setting this to be $1\%$, and taking the fairly conservative value
$\Lambda_2=400\,$MeV, I find $a_{\rm min}=0.05\,$fm.
This estimate is consistent with what CP-PACS finds in practice.

Now assume that the calculation has been improved so that $\Lambda_1=0$.
The functional form is then quadratic + cubic. Proceeding in the same way,
except this time extrapolating linearly in $a^2$,
the error is $-(a_{\rm min}\Lambda_3)^3/2$. Setting this to be $1\%$,
and taking $\Lambda_3=400\,$MeV, I find $a_{\rm min}=0.14\,$fm.
To be conservative I assume that there are
some quantities with larger discretization errors (and there are
known examples where this happens, such as $B_K$ with staggered fermions),
and so reduce the minimal lattice spacing to $0.1\,$fm.
Even so, this doubling of the minimal lattice spacing compared to the
unimproved case makes a huge difference in our ability to simulate QCD,
as discussed in sec.~\ref{sec:LQCD}.

How does one go about improving the results? Most of the work so far
has followed the Symanzik program~\cite{symanzik} 
of improvement  order by order in $a$.
Symanzik showed that, without improvement, the long-distance
($p \ll 1/a$) modes of the lattice theory can be described by an
effective continuum Lagrangian containing all operators of
dimension 5 allowed by the lattice symmetries. This is analogous to
the more familiar effective theories such as the Fermi theory of the
weak interactions. For QCD this reads
\begin{equation}
{\cal L}^{\rm eff}(a) = {\cal L}_{\rm QCD}
+ c_5(a) a \bar q \sigma_{\mu\nu} F_{\mu\nu} q + O(a^2)
\end{equation}
where $F_{\mu\nu}$ is the gluon field strength,
and $c_5$ only depends logarithmically on $a$. There are other
dimension-5 operators but these can be absorbed by a change of variables,
and so do not effect physical quantities~\cite{alpha0}.
It is the $c_5$ term which leads to the $O(a)$ corrections
in hadron masses discussed above.
To remove these corrections, one must add to the lattice action the
corresponding operator~\cite{SW}
\begin{equation}
{\cal L}_{\rm lat}^{\rm imp} = 
{\cal L}_{\rm lat}^{0} + 
	c_{\rm SW}(g^2) (a \bar q \sigma_{\mu\nu} F_{\mu\nu} q)_{\rm lat}
\end{equation}
and adjust the coefficient $c_{\rm SW}$ until $c_5$ vanishes.\footnote{%
This is not as simple as $c_{\rm SW}=-c_5$ because the addition
of the extra term leads to a shift in $c_5$ due to linear and higher
divergences in lattice integrals.} If this single tuning
can be done, then there are no linear discretization errors in any
physical quantity. Note that there are no gluonic
operators of dimension 5---they first appear at dimension 6 and give
rise to errors of $O(a^2)$. Thus for the first stage of improvement
for QCD the gluon action does not need to be improved (although it
may still be advantageous to do so).

The key, then, is to have a method for determining $c_{SW}$.
This can be done in perturbation theory, order by order in $g^2$,
and has been carried out in practice up to one-loop order.
This leads to what are called 
``perturbatively improved Wilson fermions'',\footnote{%
Other names for the improved fermion action are the
``Sheikholeslami-Wohlert'' action (``SW'' action for short)
and the ``clover'' action (based on the form of the lattice 
version of $F_{\mu\nu}$ appearing in the improvement term).}
which have discretization errors of size $\alpha_s^2 a$ and $a^2$.
It is clearly preferable, however, to have a non-perturbative determination.
A method for doing so has been developed and applied 
by the ALPHA collaboration in the last few years,
both for the quenched and two-flavor unquenched theories.
I will not discuss the method since it
was reviewed last year by Martin L\"uscher~\cite{luscher97}.
What is new this year are the results of extensive numerical tests
of ``non-perturbative $O(a)$ improvement'' as it is called. 
In other words,
with $c_{SW}$ determined non-perturbatively, do hadron masses behave
as in Eq.~\ref{eq:aextrap} with $\Lambda_1=0$?
The answer is yes!

\subsection{Numerical tests of improvement}

\begin{figure}[tb]
\psfig{figure=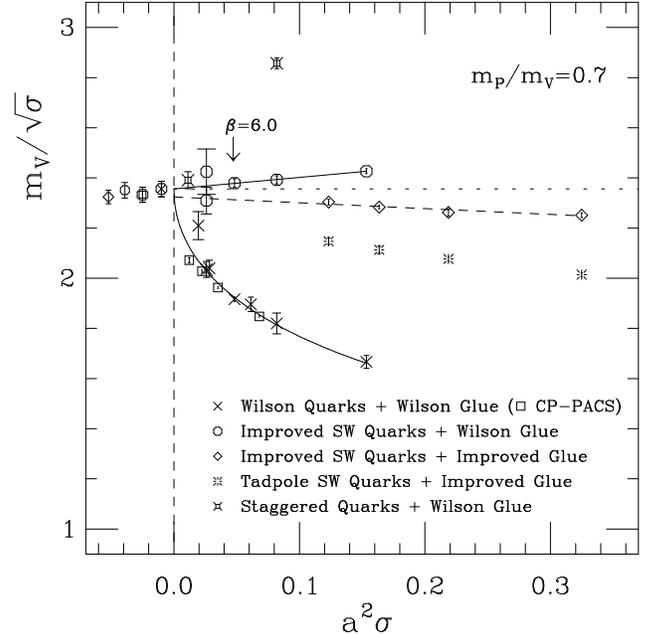,width=3.7in}
\vspace{-0.2in}
\caption{Results for $m_\rho$ versus $a^2$ at fixed $m_\pi/m_\rho=0.7$
in the quenched approximation.}
\label{fig:klassen}
\end{figure}

Studies involving several lattice spacings have been done
by the SCRI group~\cite{SCRIimp,klassenlat98}, 
the QCDSF collaboration~\cite{QCDSFimp},
and the APE-TOV collaboration~\cite{APETOVimp},
and the ALPHA collaboration~\cite{heitgerlat98}.
I will show one plot, Fig.~\ref{fig:klassen},
which nicely captures the main features of all the 
results~\cite{klassenfig}.
To test improvement one need not extrapolate  to physical light
quark masses---in fact it is better to use 
quarks with roughly the strange quark mass 
which can be simulated directly and for which the 
statistical errors are small.
This is also advantageous because it avoids so-called ``exceptional''
configurations discussed below.
The plot shows results for the vector meson mass $m_V$ versus
$a^2$ at a fixed value of the pseudoscalar to vector
meson mass ratio. The lattice spacing is determined using the
``string tension'' $\sigma$, which is the coefficient of the linear
term in the potential between a heavy quark and  antiquark in the
quenched theory. All we need to know about $\sigma$ is that it is
a gluonic quantity and thus approaches the continuum limit with
errors of $O(a^2)$. Any linear dependence on $a$ is due to
discretization errors in the fermion action.
The absolute scale is only approximately defined since this is
a quenched calculation, but the arrow in the figure labelled
``$\beta=6.0$'' roughly corresponds to $a=0.1\,$fm.

The results in the plot are from several actions involving different
degrees of improvement. The most interesting comparison is between
unimproved Wilson fermions (crosses and squares) and
non-perturbatively improved Wilson fermions (circles).
The former have errors of $O(a)$ 
while the latter should have $O(a^2)$ errors. The solid lines are the
result of fits to these hypotheses with the constraint that the
continuum value be the same. The fit is reasonable, showing
that non-perturbative improvement works.
The quantitative success is also striking.
For example, at $a\approx 0.18\,$fm the discretization errors
in hadron masses are reduced from $20-30\%$ with Wilson fermions
to $1-3\%$ with non-perturbative improvement~\cite{SCRIimp}.

If this data alone indicated the efficacy of improvement in general
then one would be able to work at larger minimal lattice spacings
than my conservative estimate of $a_{\rm min}=0.1\,$fm.
The success is not uniform, however. Heitger~\cite{heitgerlat98},
using the non-perturbatively $O(a)$ improved action,
finds that, at $a=0.1\,$fm,
the discretization errors are only a few percent in most quantities, 
but that a quantity related to $f_\rho$ differs by 28\% from
its continuum value.

The remaining points in the figure illustrate other interesting results.
One can attempt to improve the gauge action so as to reduce the $a^2$ errors
(which are large in certain glueball masses as we will see below).
This improvement has only been implemented perturbatively, reducing the
errors to $\alpha_s a^2$. One can then non-perturbatively improve
the fermions on these new background gauge fields. 
This should remove the $O(a)$ errors, but one should not expect much
reduction in the $a^2$ errors compared to the usual gauge action,
because one has not reduced the $a^2$ errors coming from the fermions.
This is what is found, as shown by
the diamonds in the figure, with the dashed line being the fit.
In this case there is no advantage gained from improving the gluon action.

The set of four points with the ``burst'' symbol (and no fit line)
show the result of perturbatively improving the fermion action.
Here one expects that there should be an $O(a)$ term,
although reduced from that for unimproved Wilson fermions
by a factor of $\alpha_s$.
Qualitatively this is what is found. 
A fit to just an $a^2$ dependence (a linear fit on the plot)
does not extrapolate close to the correct continuum value.
Note also that
the errors at $a\approx 0.2\,$fm are substantially larger
than for non-perturbatively improved fermions.

I conclude that, for almost all of the quantities studied so far,
Symanzik improvement not only removes the leading correction, but also
reduces the size of the discretization error substantially for the lattice
spacings of interest. 
This is a very encouraging result.

Staggered fermions provide an interesting exception to this rosy picture,
as shown by the two uppermost crosses in the figure. 
Staggered fermions are already improved---the errors start at $O(a^2)$---but
it appears that the coefficient $\Lambda_2$ is very large.
This is perhaps due to the fact that the effective lattice spacing is
twice as large.
Attempts to improve staggered fermions are just 
beginning~\cite{toussaint,lagae,lepagestag}---and will
be needed if they are to present a 
competitive alternative to improved Wilson fermions.

What I have discussed so far is improvement of the S-matrix: masses
and scattering amplitudes. We would also like to improve hadronic matrix
elements, such as $f_B$. This requires improvement of operators as well
as the action, e.g. for $f_B$ one needs to improve the axial current.
This is not an easy task, since for each operator there are new improvement
constants, and one wants to have a non-perturbative method for
determining all of these. So far, methods have been proposed for all
bilinear operators, and have been implemented in several 
cases~\cite{luscher97,leelat98}.
Extension to more complicated operators should be possible.

What is needed now is to extend the improved calculations towards the
chiral limit. In particular, one would like to do the improved analogue of
the CP-PACS calculation, including chiral and continuum extrapolations, 
in order to check that the results agree with the CP-PACS benchmark, 
and to determine what is the minimal lattice spacing needed.
In particular, can we use $a_{\rm min}=0.1$ or larger, 
as required to approach QCD in the near future? 
A number of calculations are underway
(as reviewed by Kenway~\cite{kenwaylat98}).
An unfortunate obstacle is the presence of exceptional configurations.
These are configurations on which the operator
$\Dslash_{\rm lat}+m$ has a zero eigenvalue for a physical quark mass,
and thus the propagator diverges at this mass. 
In the continuum these zero eigenvalues move to $m_q=0$ and
are the expected zero-modes
associated with topologically non-trivial configurations.
They are present at positive $m_q$ on the lattice because of explicit
chiral symmetry breaking.
Formally, this makes the quenched approximation ill-defined at small
quark masses with Wilson fermions~\cite{QQCDfnal}.
It turns out that this problem is exacerbated by non-perturbative
improvement, and makes it difficult to study the effect of improvement
on the quenched light hadron spectrum for lattice spacings {\em larger} than
$0.1\,$fm. This obstacle may be avoided using alternate forms of the
improved action~\cite{colorado}.
It is important to realize, however, that exceptional configurations are
quenched artifacts, and will not present
a problem in unquenched QCD because they are suppressed by the 
fermion determinant.
Thus they are a temporary obstacle.

\subsection{Other approaches to improvement}

I have concentrated so far on removing as completely as possible
the corrections linear in $a$. Other approaches to improvement
 are also being pursued. Lepage has argued for pushing Symanzik improvement
to higher order, using tadpole-improved perturbation theory 
to fix the additional improvement coefficients~\cite{lepageimp}.
Tadpole-improved perturbation theory has proven successful at predicting the
dominant lattice corrections in many cases~\cite{LM}.
The hope is to work on coarser lattices, 
perhaps as coarse as $a=0.25-0.4\,$fm, which implies $1/a=0.5-0.8\,$GeV.
For example, Alford {\em et al.}~\cite{AKLeP} argue that
errors in hadron masses could be reduced to a few percent even
at $a=0.25\,$fm if one combines tadpole estimates of the higher order
improvement coefficients with a non-perturbative determination of
the leading improvement coefficient $c_{\rm SW}$.
Their numerical results, using the D234c action, 
for which the tree-level errors begin at $a^4$, 
support this claim.

I think that this approach is very interesting, 
but I am uncomfortable at the reliance on perturbation theory
at such large lattice spacings. 
At $a=0.4\,$fm, for example, the relevant coupling constant is
$\alpha_s(\pi/a)\approx 0.4$.
This introduces two problems.
First, one is essentially giving up on doing an extrapolation in $a$, 
since several terms can contribute. 
Instead, one aims to reduce the errors to a small enough level that
no extrapolation is needed. 
Second, to estimate the residual errors requires, at the least,
calculating the full one-loop contributions to improvement coefficients 
so as to determine their effect on the final results.
It will be difficult to carry out such calculations
given the complexity of the improved action.
My major concern is for the calculation of matrix elements,
for which the perturbative uncertainty enters at leading order.
For this reason I am skeptical that one will be able, in general, to
push much beyond $a_{\rm min}=0.1\,$fm and obtain precise results for
matrix elements. Hopefully I will be proven wrong.

Another approach to improvement is to determine a good approximation
to the ``perfect action''---i.e. an action without any discretization errors.
This has been done at the classical level, but it has proven
more difficult to find a truncation appropriate for numerical simulations
which maintains the good properties of the full action.
For a recent review see Ref.~\cite{perfectrev}.

\subsection{Glueball masses in $SU(3)$ Yang-Mills theory}

The  major motivation for improvement is to make simulations of unquenched
QCD tractable. Before discussing the status of such simulations, however,
I want to give two examples from the pure gluon theory 
for which improvement has allowed substantial progress. 

If one removes not only the sea quarks but also the valence quarks from
QCD one is left with a pure gauge theory with group $SU(3)$.
Unlike quenched QCD, this is a physical quantum theory with a
unitary S-matrix. It is an interesting laboratory for lattice
methods because, like QCD, its low-energy dynamics are non-perturbative.
The spectrum  consists of interacting glueballs which, 
if massive enough, decay into multi-glueball states.
What I illustrate here is the progress that has been made 
in studying this theory through the use of improved actions.

\begin{figure}[tb]
\vspace{-0.1in}
\psfig{figure=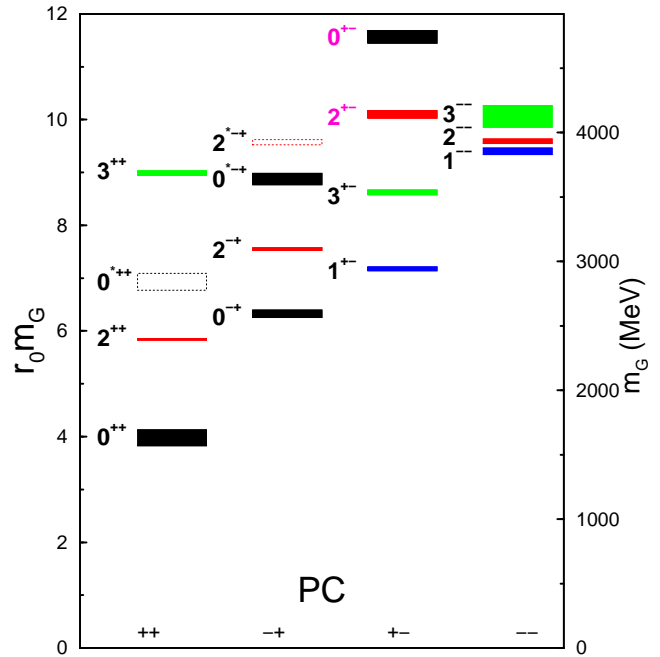,width=3.9in}
\vspace{-0.2in}
\caption{Continuum glueball spectrum in $SU(3)$ pure Yang-Mills theory.
Boxes indicate $1\sigma$ errors. Hollow boxes indicate ambiguous spin
interpretation or ambiguous continuum extrapolation.}
\label{fig:glueballs}
\end{figure}

I show in Fig.~\ref{fig:glueballs} the latest results, due to Morningstar
and Peardon~\cite{MP1}. They have obtained
the masses of the lightest 15 glueballs with small errors, 
including the extrapolation to the continuum limit.
On the LH scale,
the masses are expressed in terms of a distance scale, $r_0$, which is
determined from the quenched heavy quark potential~\cite{sommer}
(and is thus defined in the pure gauge theory). 
I stress that these results are definitive numbers for
dimensionless ratios in the pure gauge theory---there 
are no remaining approximations to be dealt with.
The effects of virtual glueball pairs are fully included.

To obtain these results two technical tricks were combined: improvement
and anisotropic lattices.
The parts of the action connecting degrees of freedom
in spatial directions were improved,
allowing the use of a large spatial lattice spacing: $a_s=0.2-0.4\,$fm.
The temporal lattice spacing was kept smaller, $a_s/a_t=3-5$,
so that the improvement of those parts of the action
involving temporal separations was not required.
This has two advantages.
First, using a smaller $a_t$ 
allows one to trace out the exponential fall-off of correlators
[$C(\tau) \propto \exp(-M \tau)$] 
more accurately and thus extract more precise values for glueball masses. 
Second, if one had to improve the action in the
temporal direction this would introduce unphysical ``ghost'' states which
would interfere with the extraction of the masses.
The use of a larger $a_s$ not only reduces the computer time required
(the calculations have been done on DEC-ALPHA workstations),
but also makes it easier to construct operators having a larger
overlap with the glueballs.

These results are a considerable advance over previous work,
which used the unimproved Wilson gauge action~\cite{UKQCDglue,GF11glue},
and needed to use lattice spacings as small as
$a=0.05\,$fm in order to extrapolate to the continuum limit.
This demonstrates the potential power of improvement:
the minimal lattice spacing is increased by a factor of four.
In particular, since the improvement was done perturbatively,
it is a test case for the adequacy of perturbative as opposed to
non-perturbative improvement.
The errors of $O(a^2)$ in the unimproved
action are reduced to $a_s^4$, $\alpha_s a_s^2$ and $a_t^2$,
and one might be concerned that the continuum extrapolations
would be difficult if all three terms were comparable.
In practice, it turns out, with one exception, that
the $a_s^4$ errors are dominant and continuum extrapolations seem
straightforward.
The exception is the scalar glueball (the lightest state),
for which it appears that the $\alpha_s a_s^2$ errors are significant,
This leads to a more difficult continuum extrapolation,
and explains why the errors for this state in Fig.~\ref{fig:glueballs}
are larger than those for most of the more massive states,
contrary to the usual pattern.
I should stress that improvement still works---the $\alpha_s a_s^2$ 
errors are smaller than the $a^2$ errors with 
unimproved Wilson fermions~\cite{MP2}.
On the other hand, if one wants to reduce the
error on the scalar glueball mass 
to a few percent using the improved action one
would have to work at smaller values of $a_s$ than $0.2\,$fm. 
This is one of the reasons why I chose the conservative
estimate $a_{\rm min}=0.1\,$fm for QCD.\footnote{%
It is possible to make alterations to the
action so as to reduce the discretization errors 
in the scalar glueball mass~\cite{MP2},
and this might lead to an action for which extrapolations could
be done using $a_{\rm min}\approx 0.2\,$fm.
While this is an interesting approach, 
it is not a substitute for a systematic improvement program
since one must make the alterations
on a case-by-case basis.}

What are the implications of these results for QCD?
The masses can be converted to physical units by matching, say,
the quenched rho mass or the heavy-quark potential to experiment.
This is done on the RH scale in Fig.~\ref{fig:glueballs}.
Let me stress, however, that the uncertainty in these predictions
will generically be larger than the 5-10\% discrepancies between
the quenched light meson and baryon masses and experiment.
This is because when we ``unquench'' glueballs, they have many
more decay channels than the light mesons and baryons, and, furthermore,
most of the glueballs can mix with quark states.
Shifts in the masses will, however, be very state dependent; for example,
they will probably be smaller
for the quark model exotics with $J^{PC}=2^{+-}$ and $0^{+-}$
which have no quark states to mix with.
Because of the potentially larger quenching errors, 
I view the results in the Figure as only a rough guide to experimental
masses,
and identification with observed glueball candidates difficult.
I should note that Weingarten and collaborators have studied the effects
of mixing and decays within the quenched approximation,
and conclude that they are small for the scalar 
glueball~\cite{weingartenglue}.
These conclusions are, however, controversial~\cite{pennington}.

Irrespective of its phenomenological implications,
the numerically determined glueball spectrum 
provides a benchmark against which to test theoretical models.
An interesting example is the prediction that states 
should be ordered roughly
according to the minimal dimension of the operators which create
them~\cite{shifmandim,jaffedim}. 
This proposal works well for the states in Fig.~\ref{fig:glueballs},
which are created by operators of dimension 4, 5 and 6.
This comparison as well as that to the bag model are
described by Kuti~\cite{kutilat98}.

\subsection{Heavy hybrids}

In pure gauge theory one can introduce an infinitely heavy quark and
antiquark pair, and calculate the energy of the induced gluonic flux as a
function of the separation $r$. 
One can excite the gluonic flux into states classified by their
angular momentum along the axis and by other discrete symmetries.
There is thus a tower of ``potentials'' which, up to an overall
constant, are physical quantities in the pure-gauge theory.

The interest in these potentials is twofold.
First, they give us insight into the physics of confinement.
In the standard picture of a flux tube of constant transverse size,
all the potentials should grow linearly at large $r$,
with the excitations corresponding to excited transverse modes.
At distances of order the transverse size, $r\approx 0.5\,$fm,
this  picture breaks down, and the detailed form of the
potentials contains information about how this happens.
The second interest is phenomenological. The potentials can
be used, in the Born-Oppenheimer approximation,
to calculate the spectrum of quarkonium states.
The lowest lying potential gives rise to the usual $Q\bar Q$ states,
while the excited potentials correspond to hybrid $Q\bar Q G$ states.
A prediction for the mass of the lightest hybrids in say, the
$b \bar b$ system would be very interesting.
Of course, such a prediction would be in the quenched approximation,
since the effects of virtual quarks (light and heavy) have been ignored.

\begin{figure}[tb]
\psfig{figure=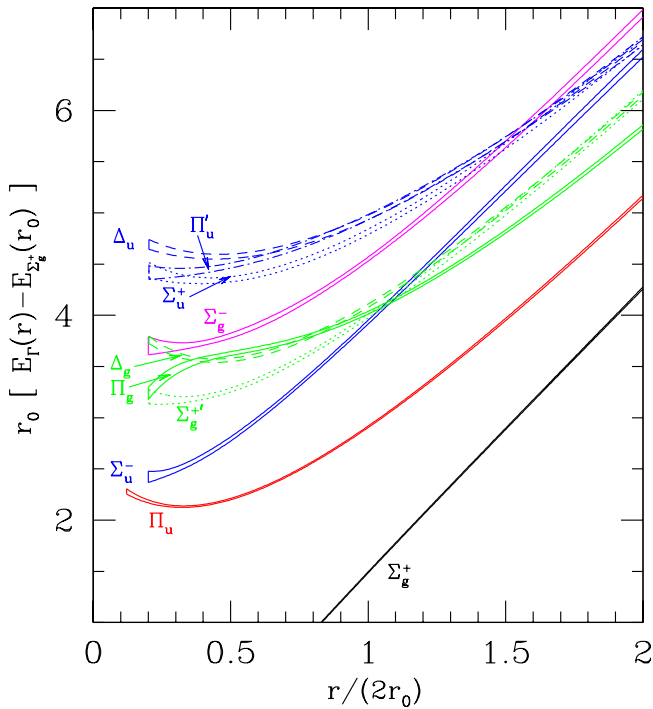,width=3.9in}
\caption{Potentials in quenched QCD after continuum extrapolation
(data of Juge {\em et al.}~\protect\cite{JKM} plotted in 
Ref.~\protect\cite{michaelrev}).
Units are $r_0\approx 0.5\,$fm. The two lines for each curve indicate
the estimated error. In the effective string theory the first four
predicted bands are: (I)$\Sigma_g$; (II) $\Pi_u$;
(III) $\Sigma'{}_g^{+}$, $\Pi_g$ and $\Delta_g$;
(IV) $\Sigma_u^\pm$, $\Pi'_u$ and $\Delta_u$.
}
\label{fig:potentials}
\end{figure}

Recent work by Juge, Kuti and Morningstar has
considerably extended our knowledge of these potentials~\cite{JKM}.
These authors use the same improved, anisotropic lattice action as in the
glueball calculation of Ref.~\cite{MP1}, 
with spatial and temporal lattice spacings
in the ranges $a_s=0.12-0.29\,$fm and $a_t=0.04-0.06\,$fm respectively.
Compared to previous work using the unimproved
action~\cite{perantonismichael}, they are able to calculate
more excited potentials and to extend the results 
out to much greater separations.
A subset of the new results are shown in Fig.~\ref{fig:potentials}.
For clarity, the potentials are shown only out to about $2\,$fm,
although they have been determined out to about $4\,$fm.
The group theoretic symbols indicate different representations
of the gluon flux,
the details of which can be found in the original work.

What I want to stress is that, as for the glueball spectrum, numerical
methods have produced detailed, accurate results for non-perturbative
quantities. These results are good enough to test theoretical models.
For example, at long distances, an effective string theory description
should apply, which predicts that the excited states should fall into
bands. This band structure is indeed observed by about $2\,$fm and 
is visible in the figure. The string model
also predicts that the differences between the levels asymptotically
approach multiples of $\pi/r$, but this turns out not to work~\cite{JKM}.
This discrepancy may be due 
to a flaw in the model itself, as discussed by Ref.~\cite{kutilat98},
or to underestimated systematic errors
(see Ref.~\cite{michaelrev} for a discussion of possibilities).
An important point, stressed by Kuti~\cite{kutilat98}, is
that the fact that the ground state potential remains very close to
linear all the way down to $r\approx 0.5\,$fm does not necessarily
provide evidence in favor of the sting picture at these distances.
Such ``precocious'' linearity is predicted by models
(e.g. the adiabatic bag model~\cite{hungarianbag})
to occur well before the onset of string-like behavior.

At intermediate distances, $r\approx 1\,$fm,
the excited states undergo a complicated pattern of level crossings.
While some of this behavior can be understood in terms of degeneracies
that are predicted for $r=0$ (Ref.~\cite{michaelrev}),
it remains a challenge to understand these levels in detail.
This work of interpretation is just beginning---for the moment the
important progress is having data good enough to compare with models.

Let me turn now to the predictions for the masses of hybrids.
In the Born-Oppenheimer approximation, the lightest $b\bar b g$
level is eightfold degenerate~\cite{griffiths}.
This level contains a $J^{PC}=1^{--}$, which could be produced in 
$e^+ e^-$ annihilation, and three exotics with
$J^{PC}=1^{-+}$, $0^{+-}$, and $2^{+-}$.
This degeneracy is lifted when one includes retardation effects,
as can be done using non-relativistic heavy quarks simulated on the
lattice. Such calculations have been done by Manke {\em et al.}~\cite{manke}
and by Juge {\em et al.}~\cite{JKM}.
The retardation effects are found to be small---for example,
all eight levels remain close in mass.
The detailed predictions from the two groups are in slight disagreement.
For example, the mass of the $b\bar b g$ hybrid with $J^{PC}=1^{-+}$ 
is found to be $11.1(1)\,$ GeV by Ref.~\cite{manke} and
$10.8\,$GeV by Ref.~\cite{kutilat98}, both in the quenched approximation.
This is an interesting mass range for hybrids, because, 
although above the $B\bar B$ threshold at $10.56\,$GeV, 
it brackets the S-wave$+$P-wave $B^{*}_J\bar B$ threshold at 
$11.01\,$GeV. There are theoretical arguments suggesting that
the dominant decay mechanism of a hybrid is to an S- and a P-wave
meson~\cite{pdg}. Thus, if the hybrid lies below or close to the
$B^{*}_J\bar B$ threshold, it could be a narrow state,
and thus observable. Improvements in the quenched results are likely
in the next year.

Calculations of $c\bar c g$ hybrids have been done
by the MILC collaboration~\cite{MILChybrids},
using improved relativistic quarks, and by
Juge {\em et al.}, using non-relativistic quarks~\cite{kutilat98}.
MILC finds $4.3 {+0.2 \atop -0.1}\,$GeV for the $1^{-+}$ state,
and the result from Juge {\em et al} is consistent with this.
           
\section{Simulations of two-flavor ``QCD''}
\label{sec:QCD}

An important step on the way to simulating full QCD is the study
of the theory with two flavors of light quarks. 
As noted in sec.~\ref{sec:LQCD}, 
the use of an improved fermion action will likely be
essential in order to simulate this theory with quarks light enough
to allow a controlled extrapolation to the chiral limit.
The hope is that improvement will work well enough to
allow a continuum extrapolation to be done
using lattice spacings no smaller than $0.1\,$fm.
The successes of improvement in the pure-gauge theory auger well
for this program.

Machines sustaining several
hundred Gigaflops have been used to do preliminary studies of two-flavor QCD. 
The extra cost of including dynamical fermions
restricts these simulations to larger lattice spacings ($a>0.1\,$fm),
smaller volumes ($L<2.4\,$fm), and larger quark masses ($m_\pi/m_\rho>0.6$)
than the state-of-the-art quenched simulations discussed above.
The most important of these restrictions is on the quark masses---reliable
extrapolations to physical quark masses are not yet possible.
Nevertheless, some interesting results have been obtained and I will
highlight a few of these.

Several collaborations have undertaken simulations aimed at studying the
spectrum of light hadrons in two-flavor QCD:
CP-PACS~\cite{burkhalter,CPPACSdyn}, 
using a perturbatively $O(a)$ improved fermion
action and an improved gauge action determined
by renormalization group methods;
MILC~\cite{MILCdyn}, using staggered fermions (which are automatically
$O(a)$ improved);
SESAM~\cite{SESAMdyn}, using unimproved Wilson fermions;
and UKQCD~\cite{UKQCDdyn}, using non-perturbatively $O(a)$ 
improved Wilson fermions.
Of these, the CP-PACS and MILC studies are the most extensive,
using several lattice spacings and the widest ranges of quark masses.

\begin{figure}[tb]
\psfig{figure=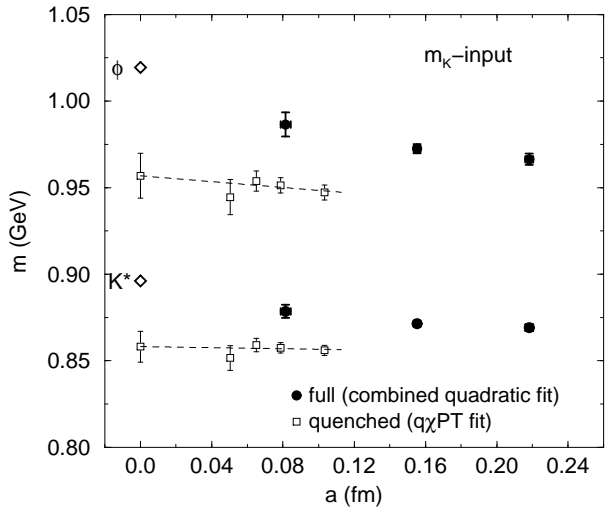,width=3.5in}
\caption{CP-PACS results for
$m_{K^*}$ and $m_\phi$ versus lattice
spacing in two-flavor QCD (filled symbols),
compared to quenched results (open symbols including continuum 
extrapolation). The strange quark mass is fixed by setting $m_K$ to its
experimental value. All results are shown after chiral extrapolation,
which is done using a polynomial fit for the two-flavor data,
and a quenched chiral perturbation theory fit for the quenched data.
The diamonds at $a=0$ show the experimental values.}
\label{fig:CPPACS-fullhyp}
\end{figure}

An interesting first question to study is whether the previously noted
discrepancy between the hyperfine splitting in quenched QCD and experiment is
reduced in two-flavor QCD. The preliminary CP-PACS results,
shown in Fig.~\ref{fig:CPPACS-fullhyp}, do show a 
significant increase in this splitting which moves the results toward the
experimental value. This is encouraging, although much more work is
needed to control the chiral and continuum extrapolations.

Other quantities for which a significant effect of unquenching has been
observed are the quark masses (discussed in the next section),
and the heavy quark potential. The latter should behave differently
both at short and at long distances. At short distances, the
one-gluon exchange contribution $\propto \alpha_s(r)/r$ should be
stronger in the unquenched theory because the coupling runs more slowly
due to the dynamical quarks. This effect has been observed 
with the expected magnitude~\cite{CPPACSstring}. 

At long distances a more striking phenomenon should occur---the string
should break! This should occur at $r\approx 1.5\,$fm, when it 
becomes energetically favorable for the flux tube to decay into two sources
screened by light (anti)quarks. At this point the potential should
bend over and reach a plateau.
Surprisingly, this rather basic phenomenon has 
not yet been seen in simulations. 
A year ago, as summarized in a review by G\"usken~\cite{gueskenLat97},
results for the potential had been extended out to distances
at which the expected flattening should begin. 
This year, CP-PACS has pushed the calculation
of the potential out to somewhat longer distances, yet still finds that
the potential keeps on rising~\cite{CPPACSstring}.
There has been progress, however, in understanding why string-breaking
is not observed~\cite{philipsen,knechtli}. 
The key point is that the operators used to create the
gluon flux have very small overlap with the broken-string state.
The authors of Refs.~\cite{philipsen,knechtli} find that if they
include in their bases operators
designed to have substantial overlap with the broken-string state, 
then they clearly observe string breaking.
These studies are done, however, on theories with scalar quarks,
and it would be very interesting to repeat them on QCD itself.\footnote{%
String breaking has already been observed in finite temperature 
QCD~\cite{laermann}. Here one has the option of using the correlation
between Polyakov lines to measure the potential and these have a
good overlap with the broken-string state as well.}

\section{Hadronic matrix elements}
\label{sec:ME}

\begin{table*}[t]
\caption{Summary of results for matrix elements. The third
column contains estimates of the effect of removing the quenched
approximation. Quark masses are quoted in the $\overline{\rm MS}$
scheme at a scale of $2\,$GeV. All $B-$parameters quoted are
renormalization group invariant, defined including non-leading order terms.
Results without references are my estimates.
}
\label{tab:ME}
\vspace{0.2cm}
\begin{center}
\begin{tabular}{|c|c|c|c|} 
\hline 
\raisebox{0pt}[20pt][10pt]{Quantity}&
\raisebox{0pt}[20pt][10pt]{Quenched Result} & 
\begin{minipage}{1in}
\begin{center}
Effect of \\
unquenching
\end{center}
\end{minipage}
&
\raisebox{0pt}[13pt][7pt]{Comments} \\
\hline\hline
\raisebox{0pt}[13pt][7pt]{$\overline m=0.5(m_u\!+\!m_d)$} & 
\raisebox{0pt}[13pt][7pt]{$4-5\,$MeV} & 
\raisebox{0pt}[13pt][7pt]{$\approx 40\%$ 
reduction~\cite{burkhalter,CPPACSdyn}} & 
\raisebox{0pt}[13pt][7pt]{
Non-perturbative matching crucial} \\
\hline
\raisebox{0pt}[13pt][7pt]{$m_s$} & 
\raisebox{0pt}[13pt][7pt]{$100-150\,$MeV} & 
\raisebox{0pt}[13pt][7pt]{$\approx 40\%$ 
reduction~\cite{burkhalter,CPPACSdyn}} & 
\raisebox{0pt}[13pt][7pt]{Using
$m_K$ and $m_\phi$ gives different results} \\
\hline
\raisebox{0pt}[13pt][7pt]{$\widehat B_K$} & 
\raisebox{0pt}[13pt][7pt]{$0.85\pm 0.06$ (Ref.~\cite{JLQCDbk})} & 
\raisebox{0pt}[13pt][7pt]{$\times 1.10\pm0.16$ (Ref.~\cite{sslat96})} &
\raisebox{0pt}[13pt][7pt]{
Unquenched result is $0.94 \pm 0.06 \pm 0.14$ 
}
\\
\hline
\raisebox{0pt}[13pt][7pt]{$f_B$} & 
\raisebox{0pt}[13pt][7pt]{$165\pm20\,$MeV (Ref.~\cite{draperlat98})} & 
\raisebox{0pt}[13pt][7pt]{${+23 \atop -0}\,$MeV (Ref.~\cite{MILCfb})} &
\raisebox{0pt}[13pt][7pt]{
Ref.~\cite{guidorev96} quotes $175\pm25\,$MeV (quenched)
}
\\
\hline
\raisebox{0pt}[13pt][7pt]{$\widehat B_B$} & 
\raisebox{0pt}[13pt][7pt]{$1.4\pm 0.1$ (Ref.~\cite{flynnsachrajda})} & 
\raisebox{0pt}[13pt][7pt]{See text} &
\\
\hline
\raisebox{0pt}[20pt][10pt]{$f_B \sqrt{\widehat B_B}$} & 
\raisebox{0pt}[13pt][7pt]{$195\pm 25\,$MeV } & 
\raisebox{0pt}[13pt][7pt]{See text } &
\raisebox{0pt}[13pt][7pt]{
Ref.~\cite{flynnsachrajda} quotes $201(42)\,$MeV (quenched)
}
\\
\hline
\raisebox{0pt}[26pt][14pt]
{$\xi=\sqrt{\mbox{$\textstyle f_{B_s}^2 \widehat B_{B_s}$}
\over \mbox{$\textstyle f_{B}^2 \widehat B_{B}$}}$} & 
\raisebox{0pt}[13pt][7pt]{$1.14{+6 \atop -5}$ (Ref.~\cite{draperlat98})} & 
\raisebox{0pt}[13pt][7pt]{See text} &
\raisebox{0pt}[13pt][7pt]{
Most reliable of all B-meson results
}
\\
\hline
\raisebox{0pt}[13pt][7pt]{$f_{D_s}$} & 
\raisebox{0pt}[13pt][7pt]{$220 {+25 \atop -20}\,$MeV (Ref.~\cite{draperlat98})} & 
\raisebox{0pt}[13pt][7pt]{${+17 \atop -1}\,$MeV (Ref.~\cite{MILCfb})} &
\raisebox{0pt}[13pt][7pt]{
Compare with experiment
}
\\
\hline
\end{tabular}
\end{center}
\end{table*}

A major reason for simulating lattice QCD is to extract
results for the hadronic matrix elements of various operators.
We would like to have both the numerical values of these matrix elements
and reliable estimates of the uncertainties in these values.
In this section I report on results for those quantities for which the
calculations are mature enough to include a continuum extrapolation
within the quenched approximation. These results are collected in
Table~\ref{tab:ME} and I will discuss them in more detail in the
following. My aim is to allow the non-specialist to make a judgement
on the reliability of the error estimates.
For more technical details,
I refer the reader to the reviews of Martinelli~\cite{guidolat98},
Gupta~\cite{rajanportugal} and Lubicz~\cite{lubiczrev}
(mainly for light quark matrix elements),
and of Flynn and Sachrajda~\cite{flynnsachrajda} and
Draper~\cite{draperlat98} (for heavy-light matrix elements).
These reviews also contain
updates on many other interesting quantities 
for which the calculations are less developed.

Three other quantities, $\alpha_s$, $m_c$ and $m_b$,
are also well determined from lattice simulations.
I do not discuss them here because there has been little new work on
them in the last year. 
Detailed discussions are given in earlier reviews:
for $\alpha_s$ see the particle data tables~\cite{pdg} and 
references therein;
for heavy quark masses see Ref.~\cite{tanmoyrajanlat97}.

\subsection{Light quark masses}

Quark masses are fundamental parameters in the standard model.
In principle, the masses of the up, down and strange quarks are encoded
in the masses of the light hadrons, but to determine them requires
a non-perturbative method. Lattice calculations will ultimately provide 
results with all errors under control, but at present they are plagued by the
uncertainties due to the quenched or partially quenched approximations.
Calculations using QCD sum rules remain for the
moment of comparable reliability, 
and competition between the two ``camps'' has produced
useful reanalyses on both sides.

Quark masses are also interesting for a more practical reason. 
In the usual parameterization of the standard model prediction
for $\epsilon'/\epsilon$ in terms of $B-$parameters, the dominant
contributions are inversely proportional to $m_s^2$.
If, as lattice calculations suggest, $m_s$ is smaller than expected a few
years ago, then the prediction for $\epsilon'/\epsilon$ is likely to increase.
Unfortunately, this is not a generic statement because there are delicate
cancelations. For recent reviews see Refs.~\cite{burasrev,ciuchinirev}.

The extraction of masses from lattice calculations is,
in principle, straightforward.
Ignoring isospin breaking, one has two dials to turn:
the bare average light quark mass $\overline m$ 
and the bare strange quark mass $m_s$. One adjusts these
until the results for two mass ratios, say $m_\pi/m_\rho$ and
$m_K/m_\pi$, agree with experiment.
One then determines the ``physical'' quark mass 
(defined in some renormalization scheme at a fixed scale)
by matching the lattice and continuum definitions of quark masses.
There are several ways of doing this---but the key point here is
that all require knowledge of matching factors relating lattice
and continuum operators.
For example, one can use the ratio of the divergence of
the axial current to the pseudoscalar density to define what is
called the ``axial Ward identity'' mass:
\begin{eqnarray}
2 \overline m^{\rm cont} 
	 &=& {\langle 0|\partial_\mu A_\mu^{\rm cont} | H\rangle \over
            \langle 0|P^{\rm cont} | H\rangle} \nonumber\\
	 &=& {Z_A \over Z_P}
	   {\langle 0|\partial_\mu A_\mu^{\rm lat} | H\rangle \over
            \langle 0|P^{\rm lat} | H\rangle} \,.
\label{eq:mbardef}
\end{eqnarray}
Here $|H\rangle$ is an arbitrary hadronic state (usually the pion), 
``cont'' and ``lat'' refer respectively to continuum and lattice regularized
operators,
and the matrix elements are to be evaluated with the
bare quark masses set to their previously determined values.
In this method the matching factors one needs are $Z_A$ and $Z_P$.
\bigskip

Let me first discuss the average of the up and down quark masses, 
$\overline m$.
The state-of-the-art a year ago was, for the most part,
to use one-loop perturbative results for matching factors,
and to do chiral extrapolations with the lowest order polynomial 
that gave a good fit.
After continuum extrapolation of all available data 
Bhattacharya and Gupta~\cite{tanmoyrajanlat97} concluded that,
in the quenched approximation,
\begin{eqnarray*}
\overline m 	&=& 4.1 \pm 0.1\, {\rm MeV} \quad ({\rm Wilson}) \\
	&=& 3.8 \pm 0.1\, {\rm MeV} 
		\quad ({\rm Perturbatively\ Improved\ Wilson}) \\
	&=& 3.5 \pm 0.1\, {\rm MeV} \quad ({\rm staggered})
\end{eqnarray*}
for the different fermion discretizations. I emphasize that these results
are for an $\overline{\rm MS}$ scale of $\mu=2\,$GeV 
(and not $1\,$GeV as often used in the sum-rule literature).
Results using different discretizations of the 
fermion action should agree in the continuum limit.
The discrepancies between them were
attributed to systematic errors associated with uncertainties
in the continuum extrapolations and matching factors.
Other groups, making different assumptions about these systematics, 
found somewhat larger values for the quark masses.

\begin{table}[tb]
\begin{center}
\caption{Comparison of perturbative and non-perturbative matching factors.
For Wilson fermions the pseudoscalar density is defined in the 
RI scheme~\protect\cite{rischeme},
while for staggered fermions the $\overline{\rm MS}$ scheme is used.
Both are defined at a scale $\mu=2\,$GeV.
All results are for quenched QCD with $\beta=6/g^2=6$ 
($a\approx 0.1\,$fm).}
\label{tab:matching}
\vspace{0.2cm}
\begin{tabular}{|c|c|c|c|} 
\hline 
\raisebox{0pt}[12pt][6pt]{Fermion type} & 
\raisebox{0pt}[12pt][6pt]{Quantity} & 
\raisebox{0pt}[12pt][6pt]{One-loop} &
\raisebox{0pt}[12pt][6pt]{Non-pert.} \\
\hline\hline
\raisebox{0pt}[12pt][6pt]{Wilson~\cite{gimenez}} & 
\raisebox{0pt}[12pt][6pt]{$Z_A$} & 
\raisebox{0pt}[12pt][6pt]{0.78} & 
\raisebox{0pt}[12pt][6pt]{0.81(1)} \\
\cline{2-4}
& 
\raisebox{0pt}[12pt][6pt]{$Z_P$} & 
\raisebox{0pt}[12pt][6pt]{0.62} & 
\raisebox{0pt}[12pt][6pt]{0.45(1)} \\ 
\hline
\begin{minipage}{1.in}
\begin{center}
{Non-pert. improved Wilson~\cite{APEimpmq}} 
\end{center}
\end{minipage}
&
\raisebox{0pt}[26pt][16pt]{$Z_P$} & 
\raisebox{0pt}[12pt][6pt]{0.59} & 
\raisebox{0pt}[12pt][6pt]{0.39(3)} \\
\hline
\raisebox{0pt}[12pt][6pt]{Staggered~\cite{ishizukalat98}} & 
\raisebox{0pt}[12pt][6pt]{$Z_A$} & 
\raisebox{0pt}[12pt][6pt]{1.0} & 
\raisebox{0pt}[12pt][6pt]{1.0} \\
\cline{2-4}
& 
\raisebox{0pt}[12pt][6pt]{$Z_P$} & 
\raisebox{0pt}[12pt][6pt]{0.54} & 
\raisebox{0pt}[12pt][6pt]{0.34} \\ 
\hline
\end{tabular}
\end{center}
\end{table}
\vspace*{3pt}

The most important improvement in the last year is 
that matching factors have been determined non-perturbatively.
The technology for doing this has been developed over the
last decade, and is an interesting example of ``applied field theory''.
The normalization of the axial current is determined by
enforcing Ward identities~\cite{boch}, 
while that of the pseudoscalar density (and other divergent operators)
is determined using off-shell quark correlation functions~\cite{npr,schr}.
I give several examples of the difference between one-loop perturbative
and non-perturbative results in Table~\ref{tab:matching}. 
The one-loop result works well for $Z_A$ but not for $Z_P$,
which is not unexpected given the sizes of the one-loop corrections in
the two cases. 
This is an example of the importance of using non-perturbative
methods to normalize operators---a point I stressed previously when
discussing the improvement program. 
Another success of non-perturbative normalization is that it
brings results for quark masses determined using different methods 
(vector versus axial Ward Identity) into agreement~\cite{rapuano}.

Referring back to Eq.~\ref{eq:mbardef}, we see that smaller values for
$Z_P$ imply larger quark masses. 
For example, the new result for staggered fermions, 
$\overline m= 4.2(3)\,$MeV~\cite{ishizukalat98},
is 20\% larger than that quoted above.
The inclusion of non-perturbative matching factors does not, however, seem to
bring the results using different types of fermion into closer agreement.
For example, there is, for the first time,
a calculation of quark masses using non-perturbatively $O(a)$ improved 
Wilson fermions~\cite{APEimpmq}, which gives $\overline m=4.9(4)\,$MeV.
Although this result does not include a continuum extrapolation
it is from a small lattice spacing, $a\approx 0.07\,$fm, for which
the $O(a^2)$ discretization errors are likely to be small. 

Another advance in the last year concerns the chiral fits.
As discussed above, there is reasonable evidence in the quenched spectrum
for the artifacts predicted by quenched chiral perturbation theory.
Fitting including these artifacts results in
quark masses that, at finite lattice spacing, are significantly reduced 
compared to those from polynomial fits~\cite{burkhalter,yoshie}. 
It turns out, however, that, with present statistics, the
shift is not significant after continuum extrapolation. 

All in all,
despite the progress that has been made, 
we are lacking the comprehensive data sets,
including non-perturbative matching factors, 
which would allow fully controlled chiral and continuum extrapolations.
My summary of the present situation is that, in the quenched approximation,
$\overline m$  lies in the range $4-5\,$MeV.\footnote{%
In the talk I mistakenly quoted a larger range $4-6\,$MeV. The upper end of
this range was based on the result from Ref.~\cite{APEimpmq}
in the RI, rather than the $\overline{\rm MS}$, scheme.}
I expect some clarification in the upcoming year, particularly using
the matching factors calculated using the Schr\"odinger functional~\cite{schr}.

The upwards movement of the quenched results brings them into reasonably
comfortable consistency with the bounds derived using QCD sum 
rules~\cite{sumrulebounds,sumruleprobs}.
But what are the quenching errors? Early calculations suggested
that quark masses were smaller when quark loops were included,
and this has been confirmed by the results from CP-PACS~\cite{CPPACSdyn}.
Their preliminary analysis suggests that two flavors of dynamical quarks
reduce the light quark masses by about 40\%, 
which might well bring them into contradiction with the sum rule bounds.
It will be very interesting to see how things evolve over the next
few years. 
There must be unexpectedly large systematic errors
in either the lattice or the sum rule methods (or both).
Suggestions for the latter have been discussed in Ref.~\cite{sumruleprobs}.
  
\bigskip

The systematic errors in determining $\overline m$ occur also
in the determination of the strange quark mass. 
In addition, as is clear from comparing
Figs.~\ref{fig:CPPACS-K} and \ref{fig:CPPACS-phi}, a different
value results if one uses the $m_K$ or $m_{\phi}$ to fix 
the strange mass.\footnote{%
There is an analogous dependence of $\overline m$ on whether 
it is determined using $m_\pi/m_\rho$ or $m_\pi/m_N$, 
but, as can be seen from Fig.~\ref{fig:CPPACS-K}, 
the difference is less statistically significant than that for $m_s$. 
It is conventional to fix $\overline m$ using $m_\rho$ rather than $m_N$,
because the former has a smaller statistical error.}
Using $m_K$ leads to $m_s/\overline m\approx 25$
if the non-linearities in $m_\pi^2/m_q$ are small, 
whereas using $m_\phi$ turns out to give 
$m_s/\overline m\approx 30$. Thus I have converted $\overline m=4-5\,$MeV
to $m_s=100-150\,$MeV for my estimate of the likely range for the
quenched strange quark mass.

\subsection{The kaon $B-$parameter}

Little has changed in the result for $B_K$ since the
reviews by Flynn~\cite{flynn96} and myself~\cite{sslat96}.
Nevertheless, since it is an important number for phenomenology,
I want to stress a few points.

The most reliable determination comes from staggered fermions,
and the best data by far is that from JLQCD~\cite{JLQCDbk}.
In a {\em tour de force} they have distinguished between
discretization errors and errors due to the use of
one-loop perturbative expressions for matching factors.
The former are proportional to $a^2$ while the latter
vary as $[\alpha_s(a)]^2\propto [\ln a]^{-2}$.
Their final result is that $B_K({\rm NDR},\mu=\,2{\rm GeV})=0.628(42)$.
The relatively large error is almost entirely due to the uncertainties
in fitting the $a$ dependence---their statistical errors
are much smaller. This is yet another indication that accurate numbers
for matrix elements require non-perturbative determinations of
matching factors.

The JLQCD data also contains another warning.
Although the calculation is automatically improved, with the
discretization errors being of $O(a^2)$, the continuum extrapolation
requires accurate data down to $a=0.05\,$fm.
This is because the dimensionful coefficients multiplying powers
of $a$ turn out to be large. In the notation of Eq.~\ref{eq:aextrap},
even though $\Lambda_1=0$, $\Lambda_2$ is a few times larger than
$\Lambda_{\rm QCD}$, as are the higher order $\Lambda_n$.
Thus improvement does not always work as expected.

Converting the JLQCD result to the renormalization group invariant quantity
$\widehat B_K$ (using the assumptions of Ref.~\cite{sslat96}) 
gives the result quoted in Table~\ref{tab:ME}.
Fortunately, as noted by Gupta~\cite{rajanportugal}, the result for
$\widehat B_K$ is quite insensitive to whether this conversion is
done using the quenched or unquenched formula.
This final quenched result differs from that quoted in 1996 because
the final JLQCD number is slightly higher than their preliminary
result due to the more sophisticated continuum extrapolation.

The estimate of quenching errors has not changed significantly
since 1996~\cite{sslat96}. 
Numerical data from unquenched simulations with two and
four flavors implies that quark loops increase $B_K$ by 
a factor of $1.05\pm 0.15$,
where the error estimate is quite conservative. 
The change from degenerate quarks to physical non-degenerate quarks in the
kaon is estimated, using chiral perturbation theory, to increase $B_K$ by
another factor of $1.05\pm0.05$.
Here the error estimate is less conservative.
Combining these two factors gives $1.10\pm 0.16$, 
where the errors have been combined in quadrature. 
Multiplying the quenched result by this factor gives the final result
quoted in the table.
Whether one should prefer this number to the quenched result,
supplemented by an estimate of the quenching error,
is a matter of taste.
I prefer to include the estimates of the effect of unquenching,
along with their fairly conservative errors, 
since they are based on reasonably reliable arguments.

\subsection{$B-$meson matrix elements}

The study of heavy-light mesons and baryons is now the major
focus of that part of the lattice community oriented towards phenomenology.
This is certainly appropriate given that B-factories 
promise a wealth of precise data in the near future.
Lattice calculations will likely play an important role
in the extraction of the underlying parameters of
the standard model from these data.
Although I report here results only for decay constants and $B-$parameters,
I want to stress that fully controlled quenched results for
many other quantities are likely to appear soon.
Examples include form factors for $B\to D^{(*)}\ell\nu$,
$B\to(\rho,\pi)\ell\nu$, $D\to K^{(*)}\ell\nu$ and $B\to K^{*}\gamma$,
and the spectrum of excited hadrons containing $b$ quarks.
For further details see the above-mentioned
reviews~\cite{flynnsachrajda,draperlat98}.

The most important theoretical issue in heavy quark calculations on
the lattice is the accuracy with which one can simulate the $b$ quark.
Direct simulation of a $b$-quark with a relativistic lattice action is
not possible because discretization errors are large
for present lattice spacings since $m_b a \gg 1$.
To avoid these errors, 
one can either use a relativistic action but extrapolate from
lighter quarks (typically somewhat heavier than the charm quark),
or treat the $b-$quark using a non-relativistic expansion.
Both approaches have potentially large systematic errors.
In the relativistic approach these are from discretization errors,
which can be reduced using improved actions and operators.
In the non-relativistic approach the errors arise from the
truncation of the non-relativistic expansion, and the need (so far at least)
to match the non-relativistic theory with QCD using perturbation theory
truncated to one-loop order. 
Although the various groups are making quite sophisticated error estimates
for their respective calculations,
the uncertainties in the errors themselves are large.~\footnote{%
For a clear discussion of the errors in the relativistic and
non-relativistic approaches see Ref.~\cite{MILCfb2} and
Ref.~\cite{tanmoylat98}, respectively.}
This makes it difficult, at least for someone not involved in
the calculations, to make a meaningful average of results from
the different groups, and so I have chosen to quote averages from
the review of Draper~\cite{draperlat98}, if available, or from
the review of Flynn and Sachrajda~\cite{flynnsachrajda}.
Fortunately, in most cases, the results from different methods are
consistent, so the central values are not likely to change much
with a different weighting of the different results.
I have illustrated this robustness 
by quoting examples of results from other reviews in the table.

\begin{figure}[tb]
\psfig{figure=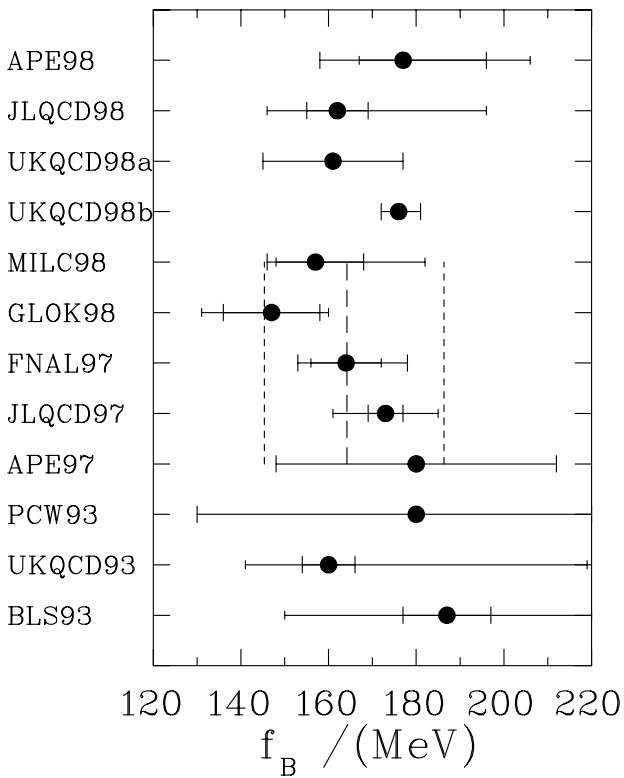,width=3.2in}
\caption{Comparison of results for $f_B$, taken from 
Draper~\protect\cite{draperlat98}. Statistical and systematic errors are
superimposed, the former indicated by the larger tick marks.
The quoted average is also shown, together with the results
included in determining it.}
\label{fig:fb}
\end{figure}

For the decays constants, Draper averages results from
the following collaborations: 
GLOK~\cite{GLOKfb}, who use non-relativistic heavy quarks; 
FNAL~\cite{FNALfb} and JLQCD~\cite{JLQCDfb}, 
who use improved Wilson quarks interpreted
non-relativistically following Ref.~\cite{EKM};
MILC~\cite{MILCfb}, who extrapolate using unimproved Wilson quarks,
interpreted following Ref.~\cite{EKM};
and APE~\cite{APE97fb} who use both unimproved and improved Wilson quarks.
The consistency between these results 
is illustrated in Fig.~\ref{fig:fb}, along with Draper's average.
Also shown are various numbers which are not included in the average:
old results which do not have as good control over systematic errors,
and new results from various collaborations, which contain
various improvements, but which are preliminary.

The quantities of particular interest to phenomenology are the
last three in Table~\ref{tab:ME}: $f_B \sqrt{\widehat B_B}$ is
needed to extract information on the CKM elements from $B-\overline B$ mixing;
$\xi$ is needed to determine the constraints from $B_s-\overline B_s$ mixing;
and $f_{D_s}$ can be compared with experimental results.
The errors in the quenched results for these quantities are
relatively small, and it is now crucial to obtain estimates
of the quenching error. For the decay constants themselves, 
such estimates have been made using numerical simulations
by the MILC collaboration, and are included in the table.
Reference~\cite{MILCfb2} gives a detailed exposition of how these
estimates are made, and notes that they are quite uncertain---
the error on the quenching error is itself estimated to be $\sim 50\%$.
For the $B$ parameters there is little numerical guidance as to the
size of quenching errors. One can make theoretical estimates,
based upon comparing chiral loops in quenched 
and full QCD~\cite{boothfb,SZfb}.
These suggest quenching errors as large as $\sim 10\%$ in
the individual $B$ parameters, 
though somewhat smaller in the ratio $B_{B_s}/B_B$.
Given the roughness of these estimates, as well as the uncertainty in the
quenching errors for the decay constants, 
it is probably premature to attempt an estimate of the unquenched results
for $f_B\sqrt{\widehat B_B}$ and $\xi$. Nevertheless, one requires
numbers in order to place bounds on CKM parameters. My best estimates
for the unquenched numbers are $f_B\sqrt{\widehat B_B}=210\pm40\,$MeV 
and $\xi=1.14(13)$. To obtain these I have included the MILC estimate for the
shift in $f_B$ due to unquenching, 
but inflated their quenching errors by 50\%, 
and taken quenching errors on $B_B$, $f_{B_s}/f_B$, and $B_{B_s}/B_B$ to be
10\%, 10\% and 5\% respectively, based on the chiral loop estimates.
Draper has given similar unquenched estimates~\cite{draperlat98}.

\section{Exactly massless quarks on the lattice}
\label{sec:chiral}

My final topic concerns discretizations of the fermion action
which exactly preserve (or almost preserve) the full global chiral symmetries 
of continuum QCD at non-zero lattice spacing.
This is to be contrasted with existing discretizations 
(Wilson and staggered fermions),
for which one cannot separate the chiral and continuum limits---chiral
symmetry is only fully restored in the continuum limit.
Separating these limits is very attractive from both a formal and
a practical point of view.
Continuum discussions of chiral symmetry breaking, topological properties,
and rigorous inequalities often rely on 
exact chiral symmetry in the massless limit.
These arguments carry over directly onto the lattice if chiral symmetry 
is maintained.
Practical calculations of quantities
constrained by chiral symmetry, such as the quark condensate and
kaon mixing and decay amplitudes 
would also be greatly simplified if there were no
explicit chiral symmetry breaking in the lattice action.
During the last year 
there has been a flurry of activity on this possibility,
and I will attempt to sketch the essence of this work
and its implications.
For more details see the recent review of Niedermayer~\cite{niedermayerlat98},
and the contribution by Neuberger~\cite{herbertichep98}.

Recall that, formally, chiral symmetry is restored in the massless limit
because the Dirac operator anticommutes with $\gamma_5$:
$\{\Dslash, \gamma_5\}=0$.
When one discretizes the fermion action, and removes the doublers
in a reasonable way (local Dirac operator with correct continuum limit),
one finds that the anticommutator no longer vanishes
\begin{equation}
\{\Dslash_{\rm lat}, \gamma_5\}=0(a)\,.
\label{eq:CSBlat}
\end{equation}
This result follows from the famous Nielsen-Ninomiya no-go 
theorem~\cite{NNnogo}.
Two lines of work have attempted to evade the
explicit chiral symmetry breaking on the RHS of Eq.~\ref{eq:CSBlat}.
The first was begun by Ginsparg and Wilson in 1982~\cite{GW}.
They noted that the $O(a)$ term in Eq.~\ref{eq:CSBlat} could, in principle, 
be made to vanish in physical matrix elements by a suitable choice
of fermion action. They formulated a condition, now known as the
``Ginsparg-Wilson relation'', the simplest form of which is
\begin{equation}
\{\Dslash_{\rm lat}, \gamma_5\}=
a \Dslash_{\rm lat} \gamma_5 \Dslash_{\rm lat}
\label{eq:GWcond}
\end{equation}
If this relation is satisfied,
then chiral Ward identities are violated only by contact terms
(because the two factors of $\Dslash$ on the rhs of Eq.~\ref{eq:GWcond}
cancel the propagators joining
the lattice artifact to other parts of correlation functions),
and thus all physical consequences of these identities 
remain valid.\footnote{%
Ginsparg and Wilson did not come upon their relation in this way,
but rather used a renormalization group argument. This 
interpretation of their result is due to 
Hasenfratz~\cite{perfectmassless}.}
In other words, the Ginsparg-Wilson relation ensures that there
is, in effect, a chiral symmetry.
Despite these nice properties,
no explicit solutions to the relation were found when gauge
interactions are included.
The difficultly is clear from an equivalent form
of the relation, 
$\{\Dslash_{\rm lat}^{-1}, \gamma_5\}= a \gamma_5$, 
which requires the anticommutator of the {\em inverse}
Dirac operator with $\gamma_5$ to be local.
Thus the Ginsparg-Wilson relation lay in obscurity for many years.

The second line of work was the attempt, by many authors over many years,
to come up with a discretized version of a chiral gauge theory
(or at least a lattice theory which one could show became a
chiral gauge theory in the continuum limit).
A very interesting approach to this problem
was initiated by Kaplan's domain wall proposal~\cite{kaplan}.
As explained in Ref.~\cite{herbertichep98}, this and other work
prompted Narayanan and Neuberger to develop the ``overlap'' formulation
of chiral gauge theories~\cite{overlap}. 
Whether this proposal works for chiral
gauge theories is a matter of some controversy, 
but it is uncontroversial that the ``overlap'' provides
a formulation of vector gauge theories such as QCD 
with exactly massless fermions.
It evades the no-go theorem by having, 
in effect, an infinite number of regulator fields.
This solution to the problem of exactly massless quarks in vector
theories also lay in relative obscurity for the last few years, 
perhaps because a practical implementation of the overlap formulation
seemed difficult.

In the last two years these threads have come together.
Neuberger showed how one could integrate out the extra
degrees of freedom in the overlap formulation and obtain a relatively
simple form for a lattice Dirac operator corresponding to
exactly massless quarks~\cite{herbertfermions}.
This he called the ``overlap-Dirac operator''.
The second step was the ``rediscovery'' of the Ginsparg-Wilson relation:
Peter Hasenfratz noted that ``classically perfect'' 
fermion actions satisfy the relation~\cite{hasenfratzlat97}.
These actions (reviewed in Ref.~\cite{perfectrev}) 
thus provide another solution to the problem of
discretizing a vector gauge theory with massless quarks.
Neuberger then noted that the overlap-Dirac operator
also satisfied the Ginsparg-Wilson relation~\cite{herbertGW}.

Finally, L\"uscher clarified the previous work by showing
that any lattice Dirac operator satisfying the Ginsparg-Wilson relation
has an exact, albeit non-local, chiral symmetry~\cite{luscherGW}.
Furthermore, he showed with a very compact argument how ``Ginsparg-Wilson
fermions'' satisfy an exact lattice version of the continuum 
Atiyah-Singer index theorem.
His work has led to further developments
in the formulation of chiral gauge 
theories~\cite{luscherabelian}.

The end result is that we now have explicit
forms of lattice fermion actions with exact chiral symmetry,
and a simple and general formalism for understanding them
based on the Ginsparg-Wilson relation.
This advance has spawned a flood both of theoretical work aimed at
determining the consequences of the Ginsparg-Wilson relation in detail,
and at finding a practical implementation.

\subsection{Numerical implementations}

My previous sketch left out what may be the most important development
from a practical point of view, namely ``domain-wall fermions''.
Kaplan's proposal for a chiral gauge theory requires an extra, fifth
dimension of infinite extent, in which the chiral theory lives on
a four-dimensional domain-wall. Truncating to a finite fifth dimension
introduces a second domain-wall inhabited by fermions of opposite
chirality, so that overall one has a vector theory.
Thus, as stressed by Shamir~\cite{shamir}, 
domain-walls provide a practical method for studying QCD.
The coupling between the left and right-handed fermions is 
suppressed exponentially by the distance between the two walls,
and so violations of chiral symmetry are controllable and can,
in principle, be made as small as one wants.
One has again succeeded in separating the chiral and continuum limits.

In the last year there have been extensive numerical simulations 
using domain-wall fermions, and they
indicate that this is a very promising approach.
In particular, it may be possible to get away with only $10-20$ sites
in the extra dimension. 
Of course, the simulations slow down by this factor, but this may
be more than compensated for by the lack of chiral symmetry breaking,
and by the fact that the domain-wall action is automatically $O(a)$
improved. For more details 
see the recent review by Blum~\cite{blumlat98}.
There have also been investigations of the properties of
the overlap-Dirac opeator and the feasibility of
numerically inverting it.
The non-locality makes inversion slow, but the potential gains are great.
I expect a lot of development in these directions during the 
upcoming year.

%

\section*{Acknowledgements}
In preparing this talk I have benefited considerably from discussions
and correspondence with
Claude Bernard, Tanmoy Bhattacharya,
Ruedi Burkhalter, Maarten Golterman, 
Steve Gottlieb, Rajan Gupta,
Shoji Hashimoto, Naruhito Ishizuka,
Julius Kuti, Peter Lepage, Paul Mackenzie, Guido Martinelli,
Bob Mawhinney, Sinead Ryan, Massimo Testa,
Doug Toussaint and Akira Ukawa. 
I particularly thank Ruedi Burkhalter, Rajan Gupta and Stefan D\"urr
for comments on the manuscript, and 
Ruedi Burkhalter for help in making the figures.
Finally, I am very grateful for the hospitality 
and support of the Center for Computational
Physics at the University of Tsukuba, Japan, 
where this write-up was completed.

\end{document}